\documentclass{aastex}

\shorttitle{Updated Nearby Galaxy Catalog}
\shortauthors{Karachentsev et al.}

\begin{document}

\title{Updated Nearby Galaxy Catalog.}
\author{Igor D. Karachentsev, Dmitry I. Makarov and Elena I. Kaisina}
\affil{Special Astrophysical Observatory RAS,  Nizhnij Arkhyz,    Karachai-Cherkessian Republic,
    Russia 369167}
\email{ikar@sao.ru}

\begin{abstract}
 We present an all-sky catalog of 869 nearby galaxies, having individual
distance estimates within 11 Mpc or corrected radial velocities
$V_{LG} < 600$ km s$^{-1}$. The catalog is a renewed and expanded version
of the ``Catalog of Neighboring Galaxies'' by Karachentsev et al. (2004).
It collects data on the following observables for the galaxies: angular
diameters, apparent magnitudes in $FUV$-, $B$-, and $K_s$- bands, $H\alpha$ and
HI fluxes, morphological types, HI-line widths, radial velocities and
distance estimates. In this Local Volume (LV) sample 108 dwarf galaxies
remain to be still without measured radial velocities.

 The catalog yields also calculated global galaxy parameters: linear Holmberg
diameter, absolute B-magnitude, surface brightness, HI-mass, stellar mass
estimated via $K$-band luminosity, HI rotational velocity corrected for
galaxy inclination, indicative mass within the Holmberg radius, and three
kinds of ``tidal index'', which quantify the local density environment.
The catalog is supplemented with the data based on the local galaxies
(http://www.sao.ru/lv/lvgdb), which presents their optical and available
$H\alpha$ images, as well as other service.

  We briefly discuss the Hubble flow within the LV, and different scaling
relations that characterize galaxy structure and global star formation in
them. We also trace the behavior of the mean stellar mass density, HI-mass
density and star formation rate density within the considered volume.

\end{abstract}

\section{Introduction}

As  previously noted by Peebles (1993), Peebles \& Nusser (2001),
Peebles et al. (2010), the study of a representative sample of the
nearest galaxies is a source of important data on
formation and evolution of large-scale structure of the
Universe. Numerous N-body simulations widely applied today in various
cosmological models suggest observational verification of their
results via comparing the properties of galaxies in a 
reference volume of fixed size.
However, almost all existing catalogs of galaxies constitute the
samples, limited by a flux (apparent magnitude), but not a
distance of galaxies. Because galaxies differ by a large range
of luminosities and surface brightnesses, the creation of a
sample, limited by a fixed volume, proves to be extremely
difficult. As an example, note that the famous Revised
Shapley-Ames Catalog (Sandage \& Tammann 1981) contains 1246
brightest galaxies  ($B_T<13.2^m$) across the sky, but only about
100 of them, i.e. 8\% make it into the sample of most nearby
galaxies with distances within 10 Mpc from us.

The study of galaxies in the LV, conditionally limited by
the $D=10$~Mpc radius, has an obvious advantage, since a lot of
dwarf galaxies, otherwise inaccessible for observations at large
distances were discovered in it. These ``test particles'' with
measured radial velocities and distances are tracing the Hubble
flow with an unprecedentedly high detail. For a comparison note
that in the most extensive Sloan Digital Sky Survey (Abazajian et
al. 2009) the average distance between galaxies with known
radial velocities is about 8 Mpc, whereas in the LV
($D<10$ Mpc), the number of galaxies with  measured velocities is
greater than 630.

Until the late 1990s development of observational cosmology in
the Local universe was hampered by the scarcity of data on 
distances of even the most nearby galaxies, located just outside
the Local Group boundaries. Deployment of unique
capabilities of the Hubble Space Telescope, combined with a new
method for determining distances to galaxies by the luminosity
of the tip of their red giant branch (TRGB)~(Lee et al. 1993) made
it possible to carry out mass distance measurements to more than
250 nearby galaxies with an accuracy of 5--10\%.
The summary of data on distances, radial velocities and other
parameters of galaxies in the LV  ($D\leq10$~Mpc) was
presented in the Catalog of Neighboring Galaxies $\equiv$ CNG
(Karachentsev et al. 2004). This volume contains  dwarf galaxies
with luminosities $10^4$ times lower than that of the Milky Way,
and it includes more than a dozen groups, similar to our Local Group
in size and population. A detailed pattern of motions of galaxies
in these groups and around them has for the first time revealed
some unexpected features in the Hubble flow at 1--3~Mpc
scales. New evidence appeared that the Hubble velocity--distance
diagrams around the Local Group and other neighboring groups are
characterized by a small dispersion of peculiar velocities
$\sim30$ km s$^{-1}$(Karachentsev et al. 2009). 
The Local Group overdensity decelerates surrounding galaxies that leads to curving the local Hubble flow.
    This effect can be observed because of small chaotic motions as well as minor distance measurement
    errors of nearby galaxies.
The achieved distance accuracy allows to determine the total mass
of nearby groups with a relative error of $\sim30$\% by the value of
radius of the ``zero-velocity sphere'' $R_0$, which separates
the group volume from remaining expanding neighborhood
(Karachentsev et al. 2009, Karachentsev 2005).

It should be emphasized that the ``$R_0$'' method gives an
estimate of the group mass, independent of the virial theorem,
and this total mass estimate refers to a scale 3.5--4.0 times larger
than the virial radius of the group. It is noteworthy that the
agreement of mass estimates of nearby groups based on external
and internal (virial) motions of galaxies is achieved only in the
presence of the cosmological parameter $\Omega_{\lambda} \simeq
0.7$.  This means that the observed properties of the Local
Hubble flow give a direct and independent evidence of the
presence in the Universe of a specific medium, the dark
energy, discovered from observations of distant Supernovae.

As shown by Dalcanton et al. (2009), Weisz et al. (2011) and other
authors,  the deep  color--magnitude diagrams obtained at the
Hubble Space Telescope  for stellar population of 
nearby galaxies provides an opportunity to reconstruct the
history of star formation in them with a resolution of
$\sim(0.1-1)$ Gyr. This approach is an important observational
tool for modeling the evolution of galaxies in different
environments.

Creation of a representative sample of galaxies in the LV
originates from a list of 179 galaxies by Kraan-Korteweg \&
Tammann (1979), which contains the galaxies with radial velocities
$V_{LG}<500$ km s$^{-1}$  relative to the Local Group centroid,
except for members of the nearby Virgo cluster. Later,
Karachentsev (1994) and Karachentsev et al. (1999) have increased
the number of galaxies in the LV to 226 and 303 objects,
respectively. In 1998--2001 Karachentseva and her colleagues
undertook a systematic search for new nearby dwarf galaxies using
the POSS-II/ESO/SERC photographic sky survey. These efforts
(Karachentseva \& Karachentsev 1998, Karachentseva et al., 1999,
Karachentseva \& Karachentsev 2000, Karachentsev et al. 2000)
along with the subsequent survey of  new objects in the HI line of
neutral hydrogen (Huchtmeier et al. 2000, 2001, 2003) have
considerably enriched the sample of galaxies in the LV. A
significant number of new irregular dwarf galaxies with radial
velocities  $V_{LG}<500$ km s$^{-1}$ were observed within the
``blind'' HI survey of the southern sky performed at the Parks
radio telescope  (Staveley-Smith et al. 1998, Kilborn et al. 2002,
Zwaan et al. 2003, Koribalski et al. 2004, Meyer et al. 2004). The
summary of these data, increasing the number of galaxies in the
LV up to N = 450 is reflected in the CNG catalog
(Karachentsev et al. 2004).

In subsequent years, the growth of the sample of the LV
occurred via detecting new dwarf galaxies within the SDSS optical
sky survey (Abazajian et al. 2009), the HIPASS (Wong et al.
2006), ALFALFA (Giovanelli et al. 2005, Haynes et al. 2011), and
Westerbork (Kova\^{c} et al. 2009)  HI surveys of the northern sky,
and as a result of systematic search for dwarf satellites of
extremely low luminosity, resolved into stars, around the Milky
Way (Willman et al. 2005, Belokurov et al. 2006), M~31 (Ibata et
al. 2007, Martin et al. 2009) and M~81 (Chiboucas et al., 2009).
By now the number of candidate members in the LV with
distances $D\leq10$ Mpc has reached 720. It is clear that
massive optical sky surveys like the Pan-STARRS (Tonry et al.
2012) and deeper ``blind'' HI surveys of the northern and
southern sky will increase this number up to 1000 and over.

\section{The Local Volume sample criterion}

Selection of galaxies in the local spherical volume of 10
Mpc radius by the condition $V_{LG}\leq500$ km s$^{-1}$, used by
Kraan-Korteweg \& Tammann (1979), assumed the Hubble parameter
value of $H_0=50$ km s$^{-1}$Mpc$^{-1}$. At the current value of
$H_0=73$ km s$^{-1}$Mpc$^{-1}$  (Spergel et al. 2007), the limit
set for radial velocities should be raised to $V_{LG}\leq
730$ km s$^{-1}$. However, the radial velocity of galaxy is
only an approximate indication of its distance. In addition to
the virial component of velocity inherent in the nearby group members, the
local velocity field is also affected by the presence of a
nearby rich Virgo cluster at a distance of 16.5 Mpc with the
velocity dispersion of $\sigma_v\simeq650$  km s$^{-1}$,
and an extensive Local Void (Tully 1988), which occupies about a
quarter of the celestial sphere. According to Tully et al.
(2008), the presence of these two main elements of the local
large-scale structure generates two velocity components of the
Local Group and surrounding galaxies:
 $\sim180$ km s$^{-1}$ towards the Virgo cluster
center  ($12^h30^m +12^{\circ}$),  and $\sim260$ km s$^{-1}$  in
the direction away from the Local Void center, located in the
region  $\sim(19^h00^m +3^{\circ}$).  An almost complete absence
of galaxies in the region of the Local Void and their relative
excess in the opposite direction creates a specific selection
effect: most of the galaxies at a distance of $D=10$ Mpc generally
have radial velocities much lower than the expected value of
$\sim730$ km s$^{-1}$.

There are recent indications that filaments and walls of the
large-scale structure may have  collective motions with an
amplitude of $\sim500$ km s$^{-1}$. Perhaps the closest example of
such great non virial motions  is the Coma~I region, where a
``flock'' of galaxies around NGC~4150 at a distance of $D\sim15$~Mpc
is moving towards us with an average peculiar velocity of $-800$
km s$^{-1}$ (Karachentsev et al. 2011). To our regret, we have to
state that the local field of peculiar velocities of galaxies is
poorly studied by now, and the proposed radial velocity correction
schemes for coherent non-Hubble motions such as the model of pure
Virgo-centric flow (Kraan-Korteweg 1986, Masters 2005) turn out to
be too simplified. Therefore, a low radial velocity of the assumed
nearby galaxy is not yet a reliable indicator of its proximity.

An ideal solution would be a direct measurement of distances to
all nearby galaxy candidates using the Hubble Space Telescope. As it
was shown by Rizzi et al. (2007), the tip of the red giant
branch method (TRGB) gives the distance accuracy of  $\sim$5\%
regardless of the galaxy morphological type. During the
exposure time with Advanced Camera for Survey at HST,
corresponding to 1--2 orbits, the TRGB
method allows to measure accurate  distances up to 7--10 Mpc,
i.e., to completely solve the problem of creating a fair sample
of the LV. A cost of the issue is, however, one to two
thousand orbits of the HST.

Other methods of  distance measurement may be either used for a
small number of objects (the Supernova method, the Cepheid
method), or be applicable only to galaxies of fixed morphology
(method of surface brightness fluctuations, Tully-Fisher
and Faber-Jackson methods), or have an accuracy not better than
25\% ( method of brightest stars).

Given all these circumstances, we have included in the LV
sample the galaxies having radial
velocities with respect to centroid of the Local Group:

$$V_{LG}<600 \,{\rm km~s}^{-1}, \eqno(1) $$
or the galaxies with distance estimates

$$D<11.0 \, {\rm Mpc}. \eqno(2)$$

A simultaneous fulfillment of both conditions (1) and (2) is not
required. Here, we took into account the fact that some
galaxies at distances   $\sim(7-10)$  Mpc may have orbital/virial
velocities that place them in the region of $V_{LG}>600$  km
s$^{-1}$ on the Hubble diagram, while other galaxies, projected
onto the Virgo cluster, are expected to
have an additional positive
velocity component due to their infall toward the
cluster.

\begin{figure*}[h]
\centerline{\includegraphics[width=0.8\textwidth,clip]{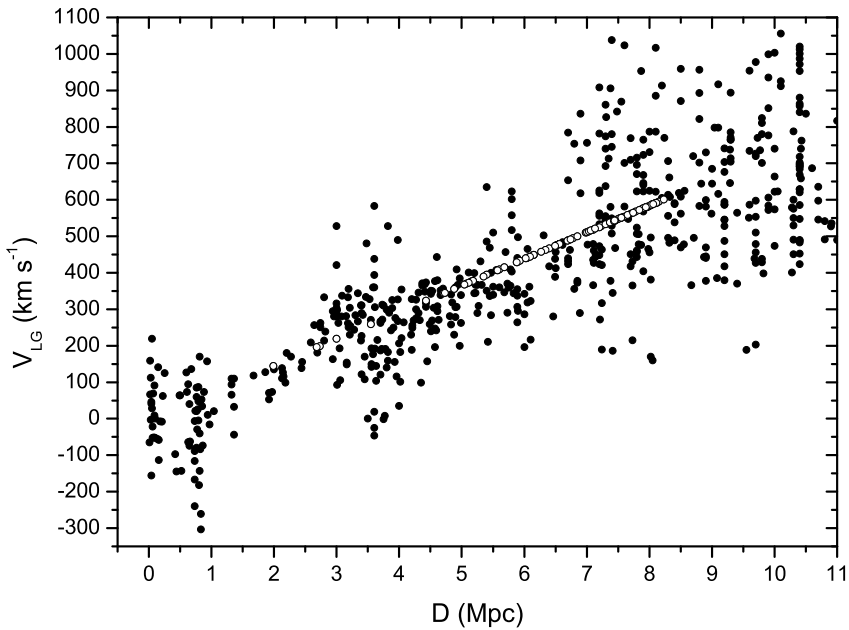}}
\caption{The Hubble flow around the Local Group centroid. Some galaxies
      in the distance range 2 -- 8 Mpc without individual distance
	estimates are drawn to trace the slope  $H_0$ = 73 km s$^{-1}$Mpc$^{-1}$ (open circles).}
\label{Fig1}
\end{figure*}

The velocity--distance diagram for 673 galaxies of the LV
is presented in Fig.~\ref{Fig1}. Beyond the upper edge of the
figure, there are 16 galaxies with $D=(7-11)$ Mpc and
$V_{LG}>1100$ km s$^{-1}$, almost all of them are located near
the line of sight, directed to the Virgo cluster center.
Beyond the right edge of the figure, 72 galaxies with
$V_{LG}<600$ km s$^{-1}$  are located, but with the distance
estimates over 11 Mpc. We did not exclude such objects from
the sample for two reasons: a) their distance estimates can not
be quite reliable, b) the distribution of such galaxies on the
sky may outline coherent motions in nearby diffuse
filaments. In addition, the list of objects in the LV
contains 108 galaxies with individual distance estimates of $D<10$
Mpc, which still have their radial velocities unmeasured.

Therefore, our total list of galaxies in the LV
consists of 869 objects. As can be seen from Fig.~\ref{Fig1}, about one
third of galaxies located in its upper right corner could be
considered to belong to the $D<10$ Mpc volume only conditionally,
since their typical distance measurement error is $\sim2$
Mpc. It also follows from these data that limiting the sample by
the condition (1) only would introduce a strong selective effect,
distorting the kinematic pattern of the LV.

\section{The replenished catalog}

Compiling the updated catalog of nearby galaxies, we generally
followed the same sequence of presenting the observational data.
Moreover, the catalog includes new parameters ($H\alpha$
emission line fluxes, far-ultraviolet fluxes) which
characterize the current star formation activity in galaxies. In
addition to this catalog, we created an online database located
at http://www.sao.ru/lv/lvgdb, which accumulates both observed and
calculated integral parameters of the galaxies, presents galaxy images,
and gives numerous references to sources of the observables.
The structure of the LV galaxy database 
is described by Kaisina et al. (2012). The list of
869 galaxies, included in the updated CNG catalog is presented in
Table~\ref{t:lv}. Its columns contain the following characteristics of
galaxies:

1. Name of the galaxy or its number in well-known catalogs.
The LV database lists all alternative  names/numbers,
including galaxy PGC number in the LEDA Extragalactic Database
(Paturel et al. 1996).

2. Equatorial coordinates of galaxy center for the
epoch (J2000.0).

3. Major angular diameter in minutes of arc, corresponding to the
Holmberg isophote ($\sim26.5^m$ arcsec$^{-2}$)  in 
B-band. The $a_{26}$ measurements  were performed visually, but 
photometric profiles of different type galaxies were used for
their calibration based on the
data by Bremnes et al. (1998, 1999, 2000) and Makarova et al.
(2009). It should be noted that
some dwarf galaxies of extremely low surface brightness
(especially those, resolved into individual stars) have the
central surface brightness fainter than the Holmberg isophote. In
these cases, the diameter $a_{26}$ rather
corresponds to the exponential scale $h$ of their brightness
profile.

4. Apparent axial ratio, measured at the Holmberg isophote.

5. Galactic extinction in the B-band according to Schlegel et al.
(1998).

6. Apparent magnitude of galaxy in far ultraviolet,
$m_{FUV},(\lambda_{eff}=1539$\AA, FWHM=269\AA) \, according to 
data on the UV-survey performed at the Galaxy Evolution Explorer
(GALEX; Martin et al. 2005, Gil de Paz et al. 2007). For 295
galaxies the asymptotic $m_{FUV}$ magnitudes were taken from Lee et
al. (2011) and are presented in Table~\ref{t:mag} without correction for
Galactic extinction. For other galaxies, the GALEX data on the
FUV-fluxes and FUV-magnitudes were extracted from the NASA
Extragalactic Database (NED), summing fluxes from all
structural knots within the optical galaxy image.

7. Apparent integral magnitude of galaxy in B-band,
sources of which are listed in the LV database as well as in Table~\ref{t:mag}. 
If a galaxy lacks the
photometric $B_T$, its apparent magnitude was estimated
by eye comparing with images of other galaxies of similar
structure with  measured $B_T$. In such cases, as a rule for
objects of low surface brightness, a typical error of $B_T$
estimate amounts to $\sim0.5^m$.

8.  Integral magnitude of galaxy in the $H\alpha$ emission line
as seen in the Cousins R- band. Following the approach by Fukugita et al. (1995),
we determined it as
$$m_{H\alpha}=-2.5\log(F_{H\alpha}) -13.64, \eqno(3)  $$
where $F_{H\alpha}$ is the integral flux in the $H\alpha$ line in
units of [erg$\cdot$cm$^{-2}$sec$^{-1}$]. The main source of
the flux data is the $H\alpha$-survey of the LV galaxies
performed at the 6-m BTA telescope of SAO RAS (Kaisin \&
Karachentsev 2006, 2008; Karachentsev \& Kaisin 2007, 2010), as
well as the survey by Kennicutt et al. (2008). 
References to $H\alpha$ data on individual galaxies are given in Table~\ref{t:mag} along with the measurement errors.
For many galaxies
their $H\alpha$ line images are presented in our LV database.

9.  Apparent magnitude of galaxy in the near infrared
$K_s$ band. The source of $K_s$ data is the 2MASS Sky Survey
(Jarrett et al. 2000, 2003) supplemented by photometric measurements
from (Fingerhut et al. 2010, Vaduvescu et al. 2005, 2006). At the lack
of accurate photometry, K-magnitudes were estimated by apparent
magnitudes in the optical (B, V, R, I) or near-infrared (J, H)
bands using synthetic color indices of galaxies from Buzzoni
(2005) and Fukugita et al. (1995). If a galaxy had only 
B-magnitude, its K-magnitude was determined by a relation
between the average color index $<B-K>$  and morphological
type (Jarrett et al. 2003): $<B-K>=4.10$  for early-types E,
S0, Sa; $<B-K>=2.35$ for late-types Sm, BCD, Ir; and
$<B-K>=4.60-0.25\cdot T$  for intermediate types $T=3-8$ by
de Vaucouleurs scale (marked by asterisk).

10. Apparent magnitude of galaxy in the HI line
$$m_{21}=17.4-2.5\log F_{HI},  \eqno(4)  $$
where the integral HI-flux is expressed in [Jy$\cdot$km s$^{-1}$].
References to data  sources on $F_{HI}$  are listed in the Table~\ref{t:mag} and in the
LV database.

11. HI line width in [km s$^{-1}$], measured at the 50\%
level from the maximum. The main source of $W_{50}$ data was 
LEDA, as well as the HIPASS and ALFALFA surveys. The references to
individual $W_{50}$ measurements are presented in Table~\ref{t:w} and the LV database.

12.  Morphological type of galaxy in the numerical code
according to the classification by de Vaucouleurs et al. (1991).
It should be noted that about three quarters of objects in the
LV are dwarf galaxies, which require a more detailed
morphological classification. For example, dwarf spheroidal
galaxies and normal ellipticals are usually denoted by the same
numerical code $T<0$, although their physical properties
drastically differ. The classification problem as well arises for
the  ``transient'' type dwarf galaxies, $Tr$, which combine the
features of spheroidal (Sph) and irregular (Ir) systems. Due to
small classification errors, such objects may ``jump'' from one
end of the  $T$- scale to the other.

\begin{figure*}[h]
\centerline{\includegraphics[width=0.8\textwidth,clip]{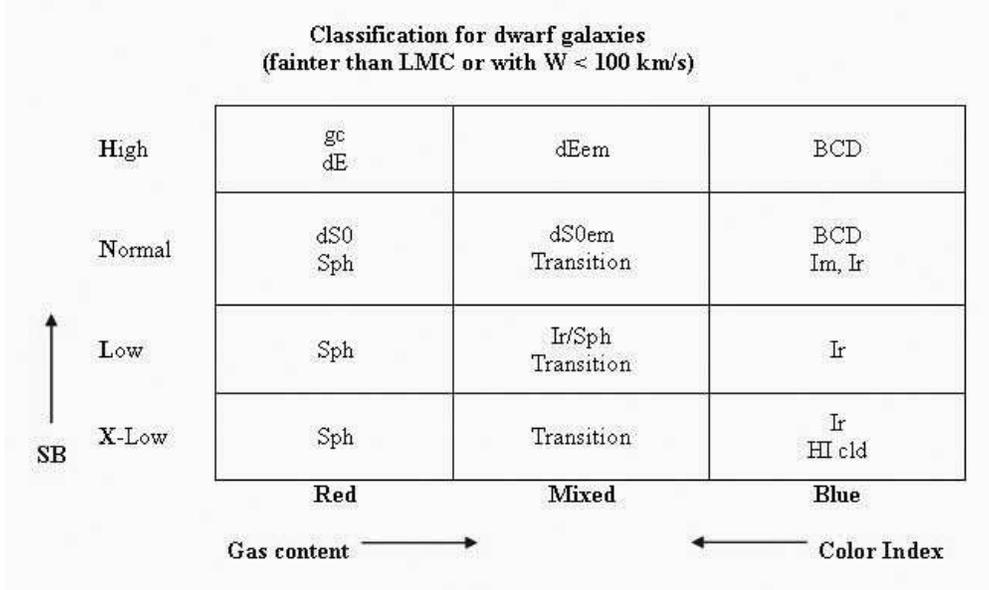}}
\caption{Classification of dwarf galaxies that takes into account galaxy
	color or gas content as well as surface brightness. The upper
	left corner potentially includes globular clusters, and the
	opposite right bottom one may contain intergalactic HI clouds.}
\label{Fig2}
\end{figure*}

13 and 14.  In order to describe the morphology of dwarf galaxies in more
detail, we introduce a two-parameter scheme, which takes
into account both the surface brightness of a dwarf system: H
--- high, N --- normal, L --- low, X --- extremely low,
 and its color (or the presence of emission):
Ir, Im, BCD - blue, Tr, dS0em, dEem --- mixed, and Sph, dE --- red.
This diagram is schematically shown in Fig~\ref{Fig2}. In its lower right
corner the intergalactic HI-clouds may be located, and
intergalactic globular clusters in the upper left. Our
classification within this scheme does not claim
for a special evolutionary significance, it rather allows to
reflect the structure of galaxies with a luminosity below that
of LMC more accurately.

15.   Heliocentric radial velocity in [km s$^{-1}$].
References to them are given in Table~\ref{t:vh}.
As a rule, we selected value of $V_h$ having the
smallest measurement error.

16 and 17. Distance to galaxy in Mpc, indicating the
method used: 
(TRGB) --- by the tip of the red giant branch; 
(Cep) --- from the Cepheid luminosity; 
(geom) --- by a geometric determination of the distance;
(SN) --- from the Supernova luminosity;
(SBF) --- from galaxy surface brightness fluctuations; 
(mem) --- from galaxy membership in known groups with measured distances of other members; 
(TF, FP) --- by the Tully-Fisher relation or by the fundamental plane; 
(BS) --- by luminosity of the brightest stars;
(CMD) --- by the color magnitude diagram using  some prominent features
or simultaneous distance and stellar population fitting;
(HB) --- by the horizontal branch;
(RR) --- from the luminosity of RR Lyrae stars;
(PNLF) --- by the planetary nebula luminosity function;
(h, h$^{\prime}$) --- by the Hubble velocity--distance relation at   $H_0$= 73 km s$^{-1}$Mpc$^{-1}$, not
accounting for (h) or in view of (h$^{\prime}$) a certain
Virgocentric flow model. In addition, we included in our sample a
small number of dwarf galaxies (N = 11), which are missing the
measurements of both optical and HI velocities due to their low
surface brightness and HI deficiency. The texture
of these objects indicates their likely proximity,
which can be verified by future observations with HST. These cases are
designated in column (17) as (txt).
References to the distance estimation are presented in Table~\ref{t:dist}.

A total list of the bibcode references is collected in Table~\ref{t:bib}.

\begin{figure*}
\centerline{
\begin{tabular}{c}
\includegraphics[width=0.8\textwidth,clip]{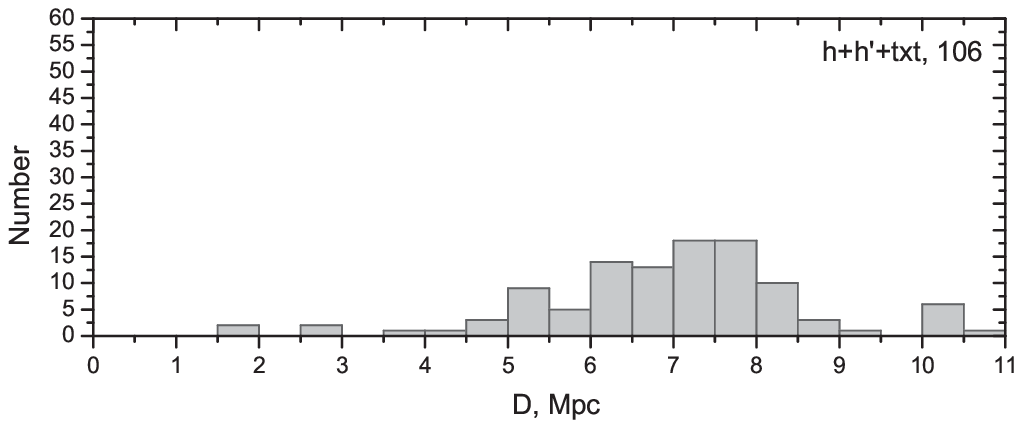} \\[-22mm]
\includegraphics[width=0.8\textwidth,clip]{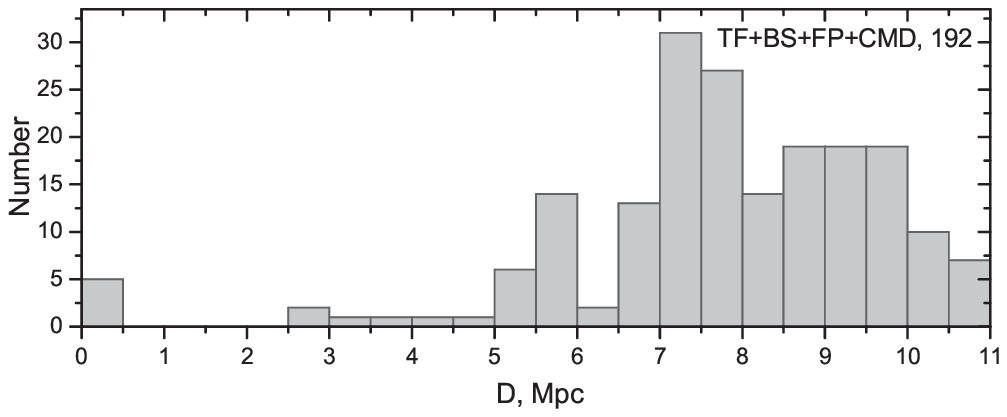} \\[-22mm]
\includegraphics[width=0.8\textwidth,clip]{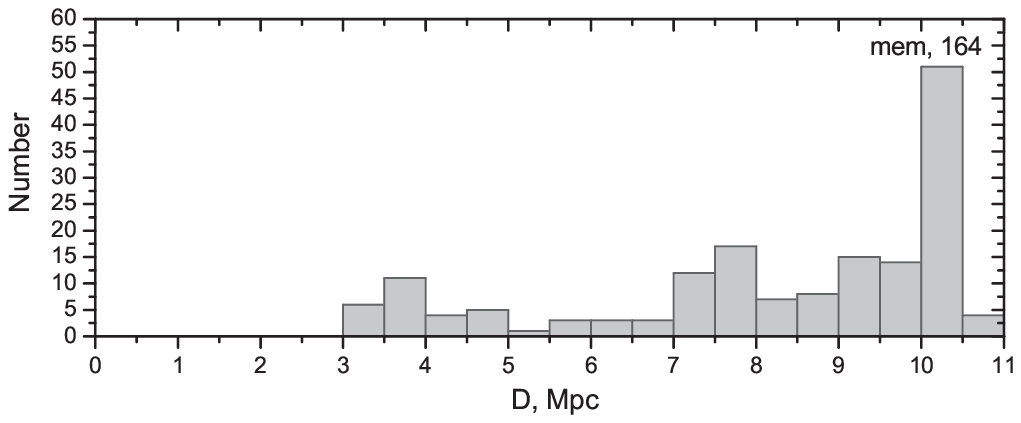} \\[-22mm]
\includegraphics[width=0.8\textwidth,clip]{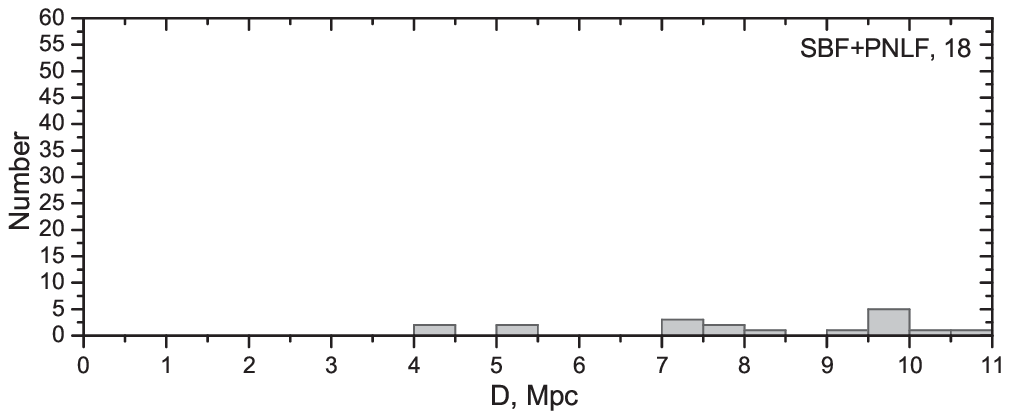} \\[-22mm]
\includegraphics[width=0.8\textwidth,clip]{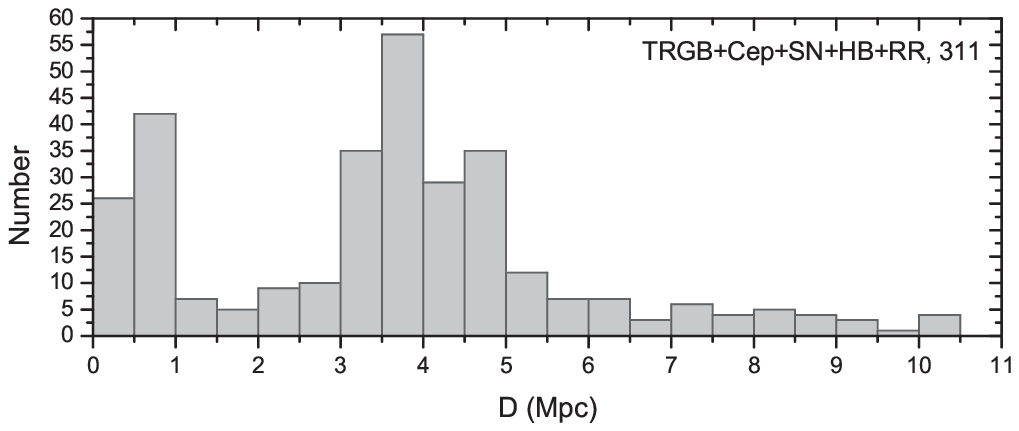} \\[-10mm]
\end{tabular}
}
\caption{Distribution of 756 nearby galaxies according to their distance
	estimates derived by different methods.}
\label{Fig3}
\end{figure*}

Fig.~\ref{Fig3} presents the distribution of galaxies on their distance
estimates, performed by various methods. For obvious
reasons, the distance medians on these panels substantially  vary
for different methods. The smallest median of $D\sim4$ Mpc falls
within the subsample  \{rgb + cep + SN\}, where the accuracy of
distance measurements is the highest, amounting to (5--10)\%.

We used the original observational data on nearby galaxies,
compiled in Table~\ref{t:lv}, to calculate their linear diameter, integral
luminosity, hydrogen mass and other global parameters.
These characteristics are presented in Table~\ref{t:param}, where
the columns contain:

1. Name of galaxy; in some cases (SDSS, APM, etc.) the coordinate part
   of name was reduced.

2. Equatorial coordinates for epoch J2000.0.

3. Major linear diameter (in kpc) at the
Holmberg isophote, corrected for Galactic extinction and 
inclination according to Fouqu\'{e} \& Paturel (1985).

4. Inclination of galaxy $i$ from the face-on position,
in degrees
$$\sin^2i=[1-(b/a)^2]\cdot[1-(b/a)^2_o]^{-1}, \eqno(5) $$
where the intrinsic axial ratio of galaxy 
($b/a)_o$ depends on its morphological type $T$ as
$$
\begin{array}{l}
\log(a/b)_o = 0.43+0.53T \,\,\,\, {\rm for} \, T\leq8 \\
\log(a/b)_o = 0.38\;\;\;\;\;\;\;\;\;\;\;\;\; {\rm for} \, T=9,10.
\end{array}
\eqno(6)
$$

%$$\log(a/b)_o=0.43+0.53T \,\,\,\, {\rm for} \, T\leq8$$
%$$ {\rm and} \eqno(6)$$
%$$\log(a/b)_o=0.38\,\,\,\,\,\,\, {\rm for} \, T=9,10. $$

We adopted this expression from Paturel et al. (1997) with a correction
for slight difference between our type scale and the LEDA
$T$-scale. Therefore, for the Ir, Im, BCD-type galaxies we assume
the intrinsic axial ratio to be  ($b/a)_o=0.42$ in accordance with 
statistics of apparent axial ratios of these galaxies.

5. Amplitude of rotational velocity of galaxy
$V_m=W^c_{50}/2\sin i$, adjusted for the inclination, where the
HI line width, $W^c_{50}$, contains a correction for
turbulent motions following Tully \& Fouqu\'{e} (1985) scheme with
a parameter $\sigma_z=10$ km s$^{-1}$.

6. Internal extinction in galaxy in the B-band
according to Verheijen, 2001:
$$ A^i_B=[1.54+2.54(\log 2V_m-2.5)]\log(a/b), \eqno(7)$$
if $2V_m>78$ km s$^{-1}$, otherwise $ A^i_B=0$.  Therefore,
dwarf galaxies with $V_m<39$ km s$^{-1}$, as well as gas-poor
E, S0-galaxies without $W_{50}$ estimates, were considered to be
fully transparent systems.

7. Absolute B-band magnitude of galaxy, corrected for the
Galactic and internal extinction.

8. Average surface brightness in the B-band within the
Holmberg isophote (mag arcsec$^{-2}$)
$$ SB_B=B^c_T+5\log a_{26}^c+8.63, \eqno(8) $$
where the apparent magnitude and angular diameter are corrected
for extinction and inclination.

9. Logarithm of $K_s$-band luminosity of galaxy in solar units
corrected for extinction $K-K^c=0.085(A^G_B+A^i_B)$, at 
absolute magnitude of the Sun $M_K^{\odot}=3.28$ (Binney \&Merrifield 1998).

10. Logarithm of indicative  mass $M_{26}$ within the Holmberg
radius, where the mass
$$M_{26}=3.31\cdot10^4V^2_m\cdot a^c_{26}\cdot D \eqno(9)$$ is expressed
in solar masses, corrected angular diameter in minutes of arc, $V_m$ in  [km s$^{-1}$], and $D$
--- in [Mpc] (Roberts \& Haynes 1994).

11. Logarithm of hydrogen mass
$$M_{HI}=2.356\cdot 10^5\cdot D^2\cdot F_{HI}, \eqno(10)$$
where $M_{HI}$ is expressed in solar masses, $D$ --- in Mpc, and
$F_{HI}$ --- in [Jy km s$^{-1}$] (Roberts \& Haynes 1994).

12.  Radial velocity of galaxy relative to the 
Local Group centroid with apex parameters (Karachentsev \& Makarov 1996) adopted in NED:
$$V_{LG}=V_h+316[\sin b\cdot\sin(-4^{\circ})+\cos b\cdot\cos(-4^{\circ})\cdot\cos(l-93^{\circ})], \eqno(11)$$
where $l$ and $b$ are Galactic coordinates of the galaxy.

13. Tidal index $\Theta_1 $, determined below in Section~8
via distance and mass of the nearest significant neighbor. Positive
values of $\Theta_1$ indicate the membership in groups, negative
values correspond to isolated galaxies.

14. Name of the ``main disturber''(=MD), i.e. the neighboring galaxy,
producing the maximal tidal influence on this galaxy. In fact, the
set of objects in Table~\ref{t:param} with the same MD and $\Theta_1>0$
corresponds to a definition of physical group of galaxies.

15. Another tidal index (or density contrast), determined by five most
important neighbors. Unlike $\Theta_1$, the index
$\Theta_5$ seems to be a more robust feature of galaxy environment.

16. Logarithm of the K-band luminosity density within 1 Mpc around the galaxy,
taken in units of the mean global $L_K$ density derived
from 2MASS by Jones et al. 2006.

\section{On the sample completeness}

Degree of completeness in the number of galaxies presented in
a sphere of 10 Mpc radius can be estimated only approximately,
since it depends on many factors which are hard to verify. The
optical and HI surveys of the sky are usually limited to a fixed
flux, but the difference of galaxies in luminosity, size, surface
brightness and gas abundance is enormous. High-accuracy distances
to most of the galaxies are as yet known within $\sim5$ Mpc. Given
the distance measurement errors by the Tully-Fisher method of
$\sim(20-25)$\%,  a significant part of   galaxies at the
periphery of the LV are only its conditional members.
The presence of collective non-Hubble motions on the scale of
$\sim10$ Mpc also makes it difficult to assess the completeness of
our sample, since the amplitude of these motions ($\sim300$ km
s$^{-1}$) may make up a half of the velocity constraint (1) we
have adopted. Note that the ``Zone of Avoidance'' behind the Milky Way
is already well filled with nearby galaxies owing to the
systematic surveys in the HI line.

\begin{figure*}

\centerline{\includegraphics[width=0.8\textwidth,clip]{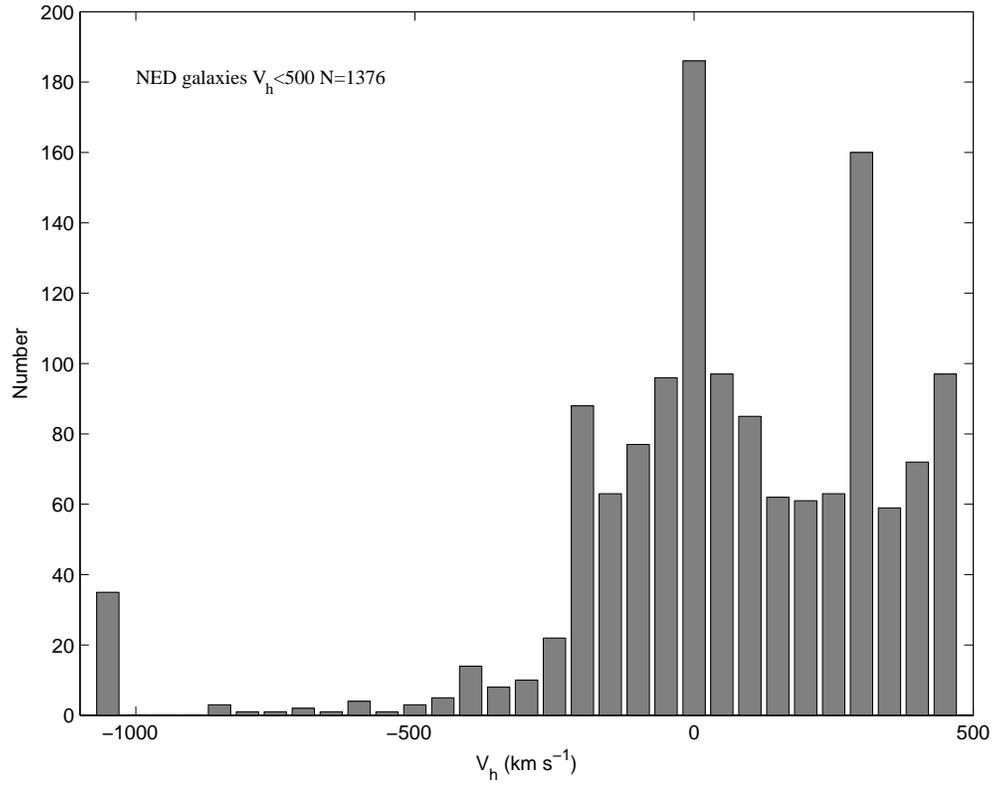}}
\caption{Number vs. Heliocentric velocity distribution for 1376 non-galaxy objects
	of low velocities from automatic surveys indicated in NED as
	``galaxies''.}
\label{Fig4}
\end{figure*}

We should note that the evaluation of completeness of the LV
sample is also affected by a problem of ``astro-spam''.
Massive automatic surveys of galaxy redshifts produce a
significant number of cases where the radial velocity of a star,
projected onto a distant galaxy is falsely attributed to this
galaxy. For example, the DEEP2 spectral survey
(http://deep.berkeley.edu) and other recent surveys give more than
1000 fictitious objects (see Fig.~\ref{Fig4}) included in NED as ``galaxies'' with
velocities  $V_{h}<500$ km s$^{-1}$. This implies that the amount
of ``rubbish'' from the automatic surveys is greater than the
actual population of the LV. Another source of
astro-spam is the confusion at the optical identification of HI
sources in surveys with a low angular resolution. The
presence around our Galaxy of hundreds of high velocity clouds
with low line-of-sight velocities and small  $W_{50}$ widths also
provokes the inclusion of false ``nearby'' dwarf galaxies in the
LV (plenty of such examples can be found in the paper
by C\^{o}t\'{e} et al. 1997). Finally, galaxy databases account for very
exotic cases, for instance, a ``galaxy'' AM~0912-241 (see NED)
with radial velocity of  +614 km s$^{-1}$ (Mathews et al.
1995) which is in fact only a photographic emulsion defect.

\begin{figure*}

\centerline{
\begin{tabular}{c}
\includegraphics[width=0.8\textwidth,clip]{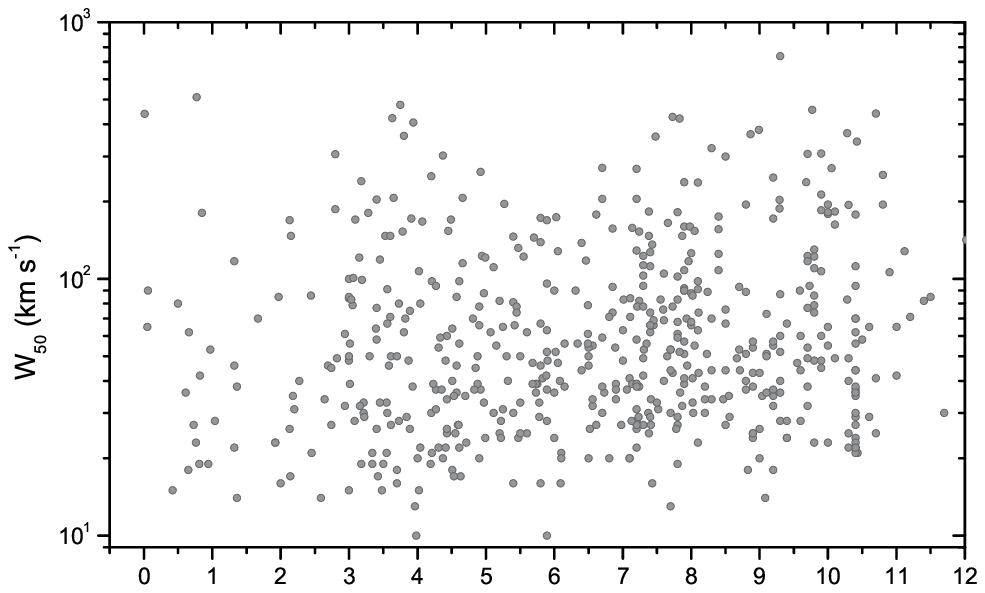} \\[-22mm]
\includegraphics[width=0.8\textwidth,clip]{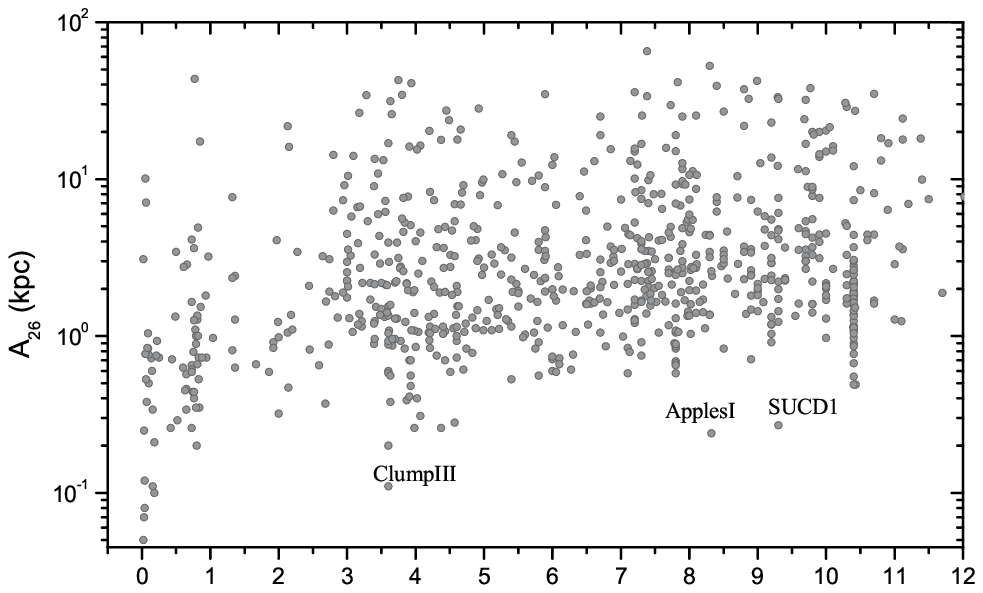} \\[-22mm]
\includegraphics[width=0.8\textwidth,clip]{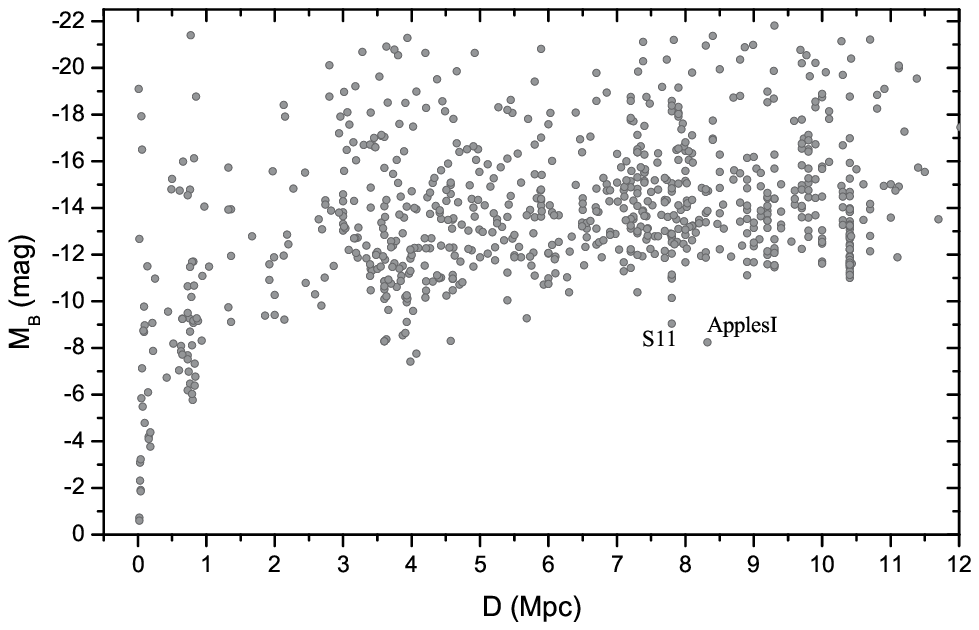} \\[-10mm]
\end{tabular}
}
\caption{Distribution of nearby galaxies according to their absolute
	magnitude (bottom), linear Holmberg diameter (middle) and HI line
	width (top). Some ultra faint dwarfs are denoted.}
\label{Fig5}
\end{figure*}

Some idea of the  completeness rate of the LV sample is given
in Fig.~\ref{Fig5}, where the panels demonstrate the distributions of
galaxies by absolute B-magnitude, linear diameter $A_{26}$ and
line width $W_{50}$ depending on distance. These data imply that
the galaxies with absolute magnitudes brighter than $-11^m$ or
linear diameters  $A_{26} > 1.0$ kpc occur as frequently both in
the nearby and distant parts of our sample. In this sense, the
conditional level of completeness of the LV sample is about
(40-60)\%, which is indirectly confirmed by the distribution of
$W_{50}$ vs $D$. However, among the members of the Local Group
($D<1$ Mpc), only a half of the galaxies have absolute magnitudes
brighter than $-11^m$. Consequently, more than a half of ultra faint dwarf
companions around normal galaxies, like the Sombrero galaxy
($D=9.3$ Mpc), still remain outside our field of view.

A case of an isolated dwarf spheroidal galaxy Apples~I with
$M^c_B=-8.30^m$ and $A_{26}^c=0.25$ kpc at a distance of 8.5 Mpc
is intriguing here. This galaxy was accidentally discovered by
Pasquali et al. (2005) as a system resolved into stars with HST.
The statistics of images obtained so far with HST shows
(Karachentsev et al. 2009) that the population of such 
difficult-to-detect tiny crumbs without gas and with only old
stellar population may reach up to $N\sim10^3-10^4$ in the
LV, i.e. be dominant over other types of galaxies.

\begin{figure*}

\centerline{
\begin{tabular}{c}
\includegraphics[width=0.8\textwidth,clip]{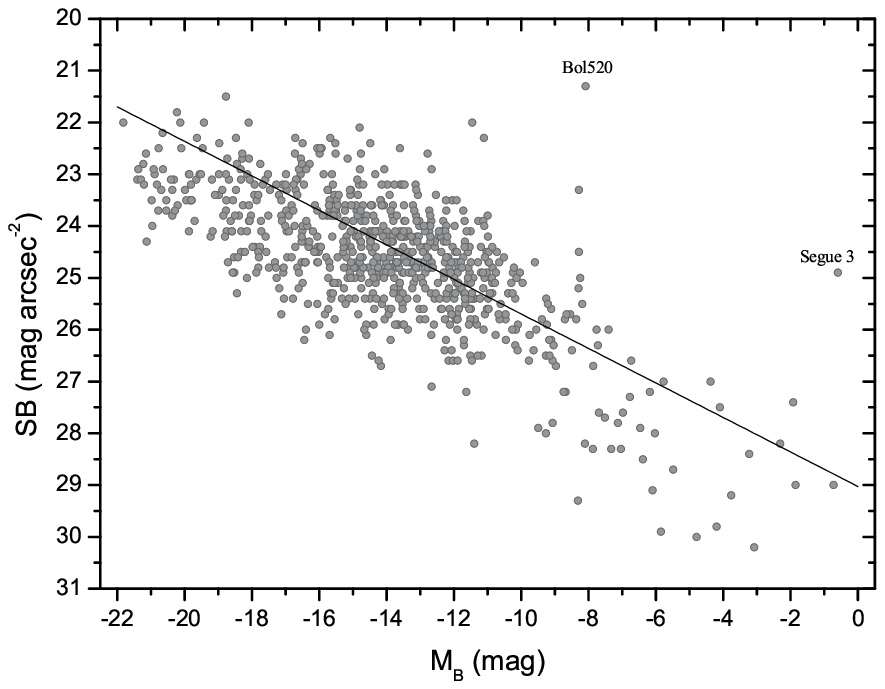} \\[-22mm]
\includegraphics[width=0.8\textwidth,clip]{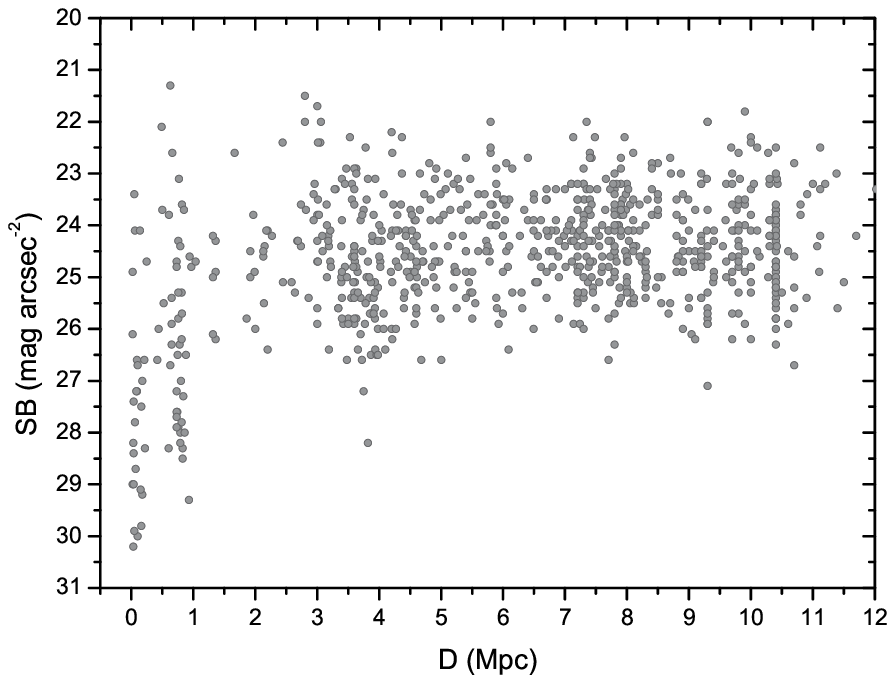} \\[-10mm]
\end{tabular}
}
\caption{Mean surface brightness of the LV galaxies versus their absolute
	magnitude (bottom) and distance (top). The line traces a case of
	constant spatial luminosity density within the Holmberg radius.}
\label{Fig6}
\end{figure*}

The upper panel in Fig.~\ref{Fig6} shows the distribution of mean surface
brightness of the LV galaxies at different distances. If we
exclude the dwarf systems of the Local Group with SB$>$26.5 mag
arcsec$^{-2}$, then the mean surface brightness is almost
independent of distance. Dwarf companions of the Andromeda and
Milky Way with  SB=(27--31) mag arcsec$^{-2}$ are distinguishable
at close distances only due to the fact that they are resolved
into individual stars. It is obvious that a large number of such
satellites can also exist around more distant normal galaxies. A
special survey of similar objects in a region of the M~81 group
(Chiboucas et al. 2009) confirms this assumption.

In cosmological models, describing the structure of dark galaxy halos,
the central density of matter is presumed to be the same
regardless of size and mass of the halo (Navarro et al.
1996). At that, the mass and size of the halos are related as
$M_{DM}\sim R^3$. If the distribution of stellar matter follows
the distribution of dark matter, we can expect a similar relation
$L\sim A^3$ between the integral luminosity of galaxy and its
diameter, which implies the relation SB$\sim(1/3)M_B$. Such a
relationship is indicated at the bottom plot of Fig.~\ref{Fig6} by
straight line. As one can see, the LV sample as a whole follows
the expected relation fairly well. The deviation from it at the
extremely low surface brightness end is due to a systematic
overestimation of dwarf galaxy sizes, the brightness profiles of
which entirely lie  below the Holmberg isophote.

\section{Galaxy distribution within 11 Mpc}

\begin{figure*}

\centerline{
\begin{tabular}{c}
\includegraphics[width=1.2\textwidth,clip]{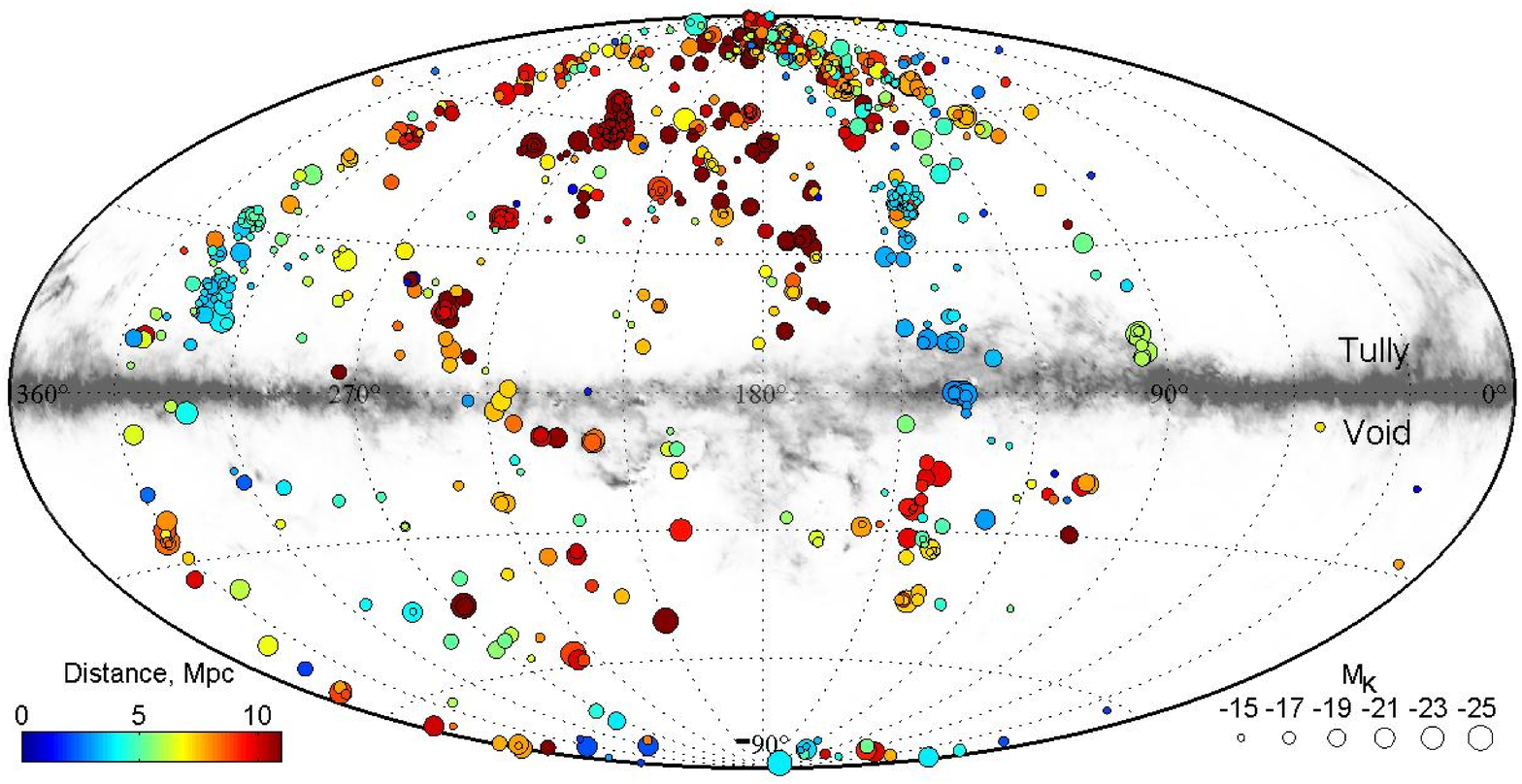} \\
\includegraphics[width=1.2\textwidth,clip]{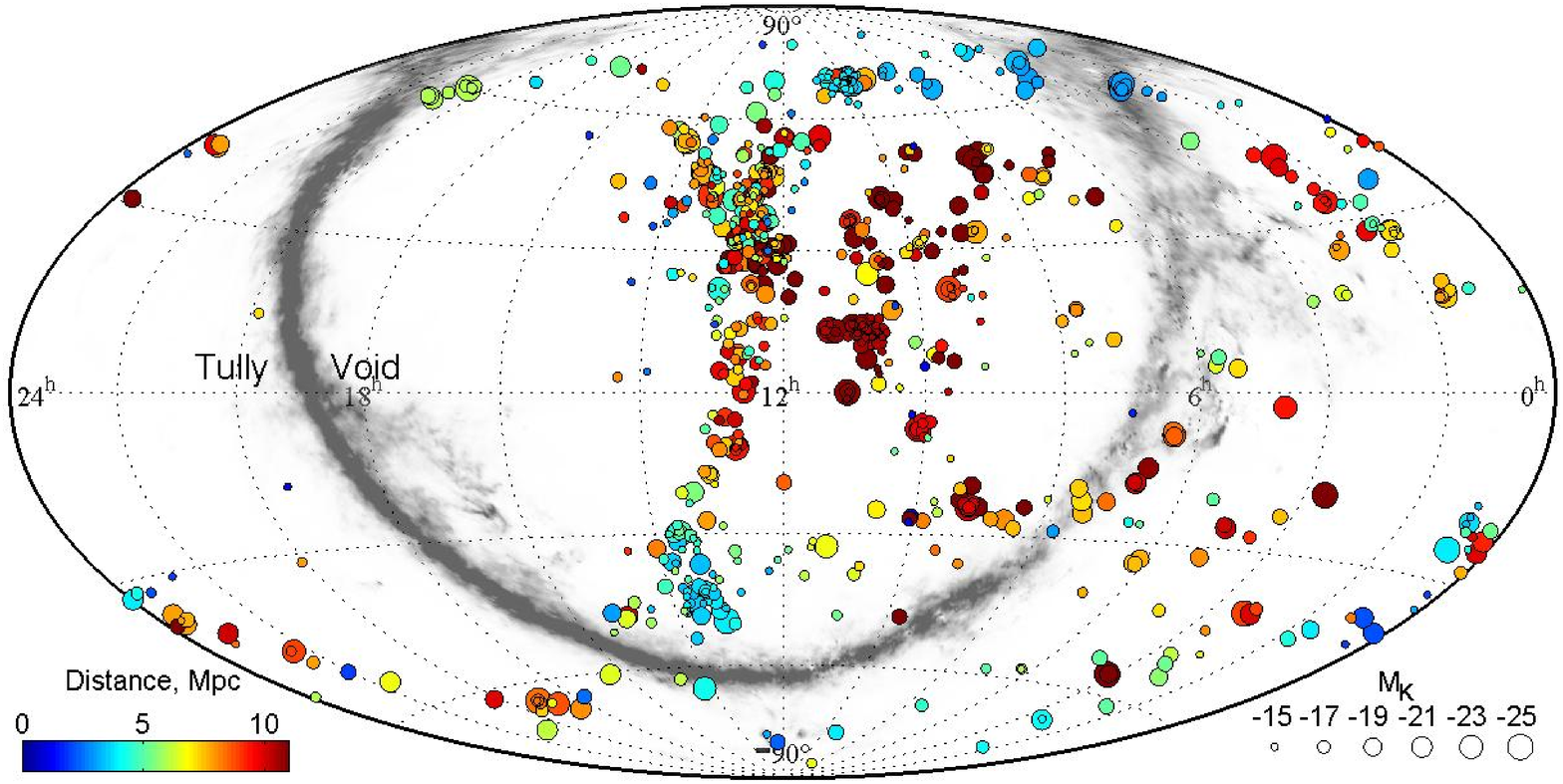} \\
\end{tabular}
}
\caption{Distribution of nearby galaxies on the sky in equatorial (bottom)
	and galactic (top) coordinates. The Local Group members are not
	shown. Galaxy distance and luminosity are indicated by circles
	of different color and size. The Zone of Avoidance in the Milky
	Way is outlined by gray lane.}
\label{Fig7}
\end{figure*}

Distribution of the LV galaxies on the sky  is
shown in Fig.~\ref{Fig7} in the equatorial (bottom panel) and galactic
(top panel) coordinates. The galaxies with distances of
$D=(1-11)$ Mpc are shown by circles, whose size reflects the
luminosity of galaxy, and color --- its distance. Here the
members of the Local Group with $D<1$ Mpc were excluded. The
shaggy gray lane in the panels corresponds to the region of strong
extinction in our Galaxy. The distribution of galaxies shows their
concentration in the region of well-known nearby groups around
M~81, Centaurus~A, M~83, IC~342, NGC~253, M~101, NGC~6946, Leo~I,
etc. The figure exhibits an extensive area in the Hercules-Aquila,
almost completely devoid of galaxies, i.e. the Local Void (Tully
1988). This void extends far beyond the LV boundary.

\begin{figure*}

\centerline{\includegraphics[width=0.6\textwidth,clip]{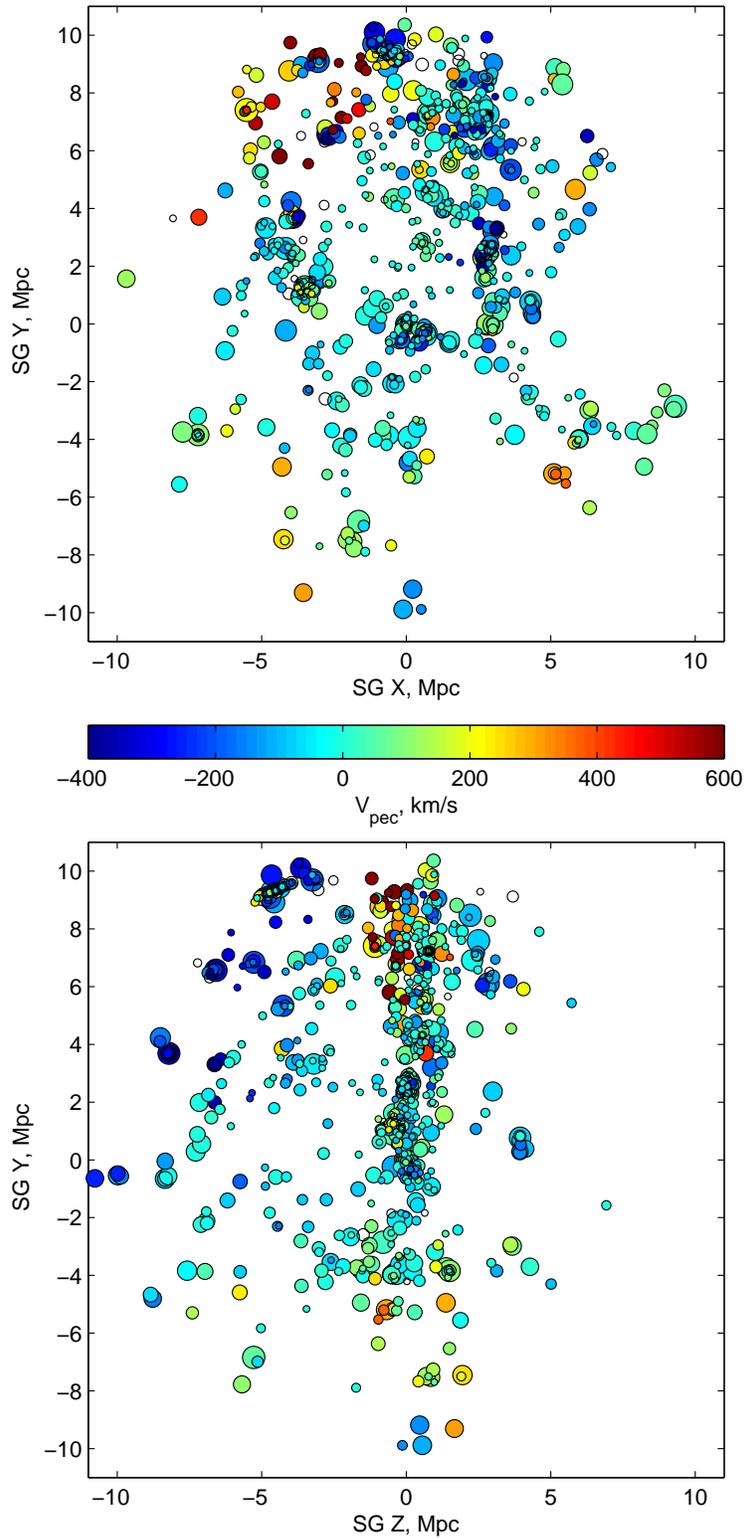}}
\caption{Distribution of the LV galaxies within 11 Mpc in Cartesian Super-
	galactic coordinates. Luminosity of the galaxies is marked by a size
	of circles and peculiar velocities are indicated by the color scale.}
\label{Fig8}
\end{figure*}

Distribution of galaxies within a sphere of 11 Mpc radius
is shown in the panels of Fig.~\ref{Fig8} in different projections of
Cartesian Supergalactic coordinates SGX SGY, SGZ. The size
of circles characterizes luminosity of galaxies and their
color marks peculiar velocity  $V_{pec}=(V_{LG}-73\cdot D)$
according to the scale between the panels. The clustering of
galaxies in groups, as well as towards the Supergalactic plane is
clearly noticeable. These panels show that the galaxies with
high negative peculiar velocities are located far away from the
Supergalactic plane, and the galaxies with $V_{pec}> +300$  km
s$^{-1}$ are concentrated in the region of SGY $>6$ Mpc, showing
a tendency of infall towards the Virgo cluster (its coordinates
are SGY = +16 Mpc, SGX $\simeq$ SZY $\simeq$ 0). These data give
us an idea that the pattern of motions in the LV is
rather far from the unperturbed Hubble flow.

\section{Morphology and optical scaling relations}

\begin{figure*}

\centerline{
\begin{tabular}{c}
\includegraphics[width=0.8\textwidth,clip]{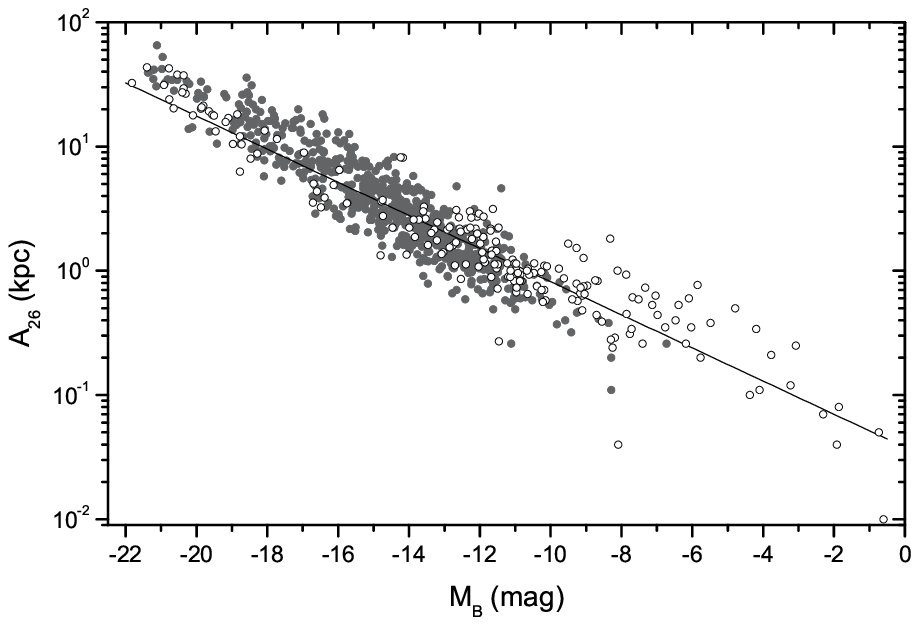} \\[-23mm]
\includegraphics[width=0.8\textwidth,clip]{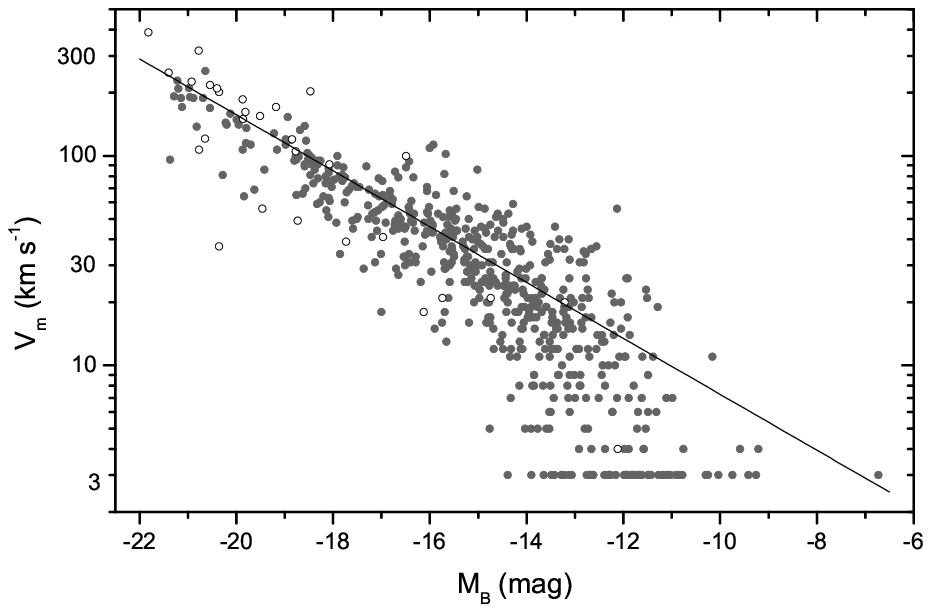} \\[-23mm]
\includegraphics[width=0.8\textwidth,clip]{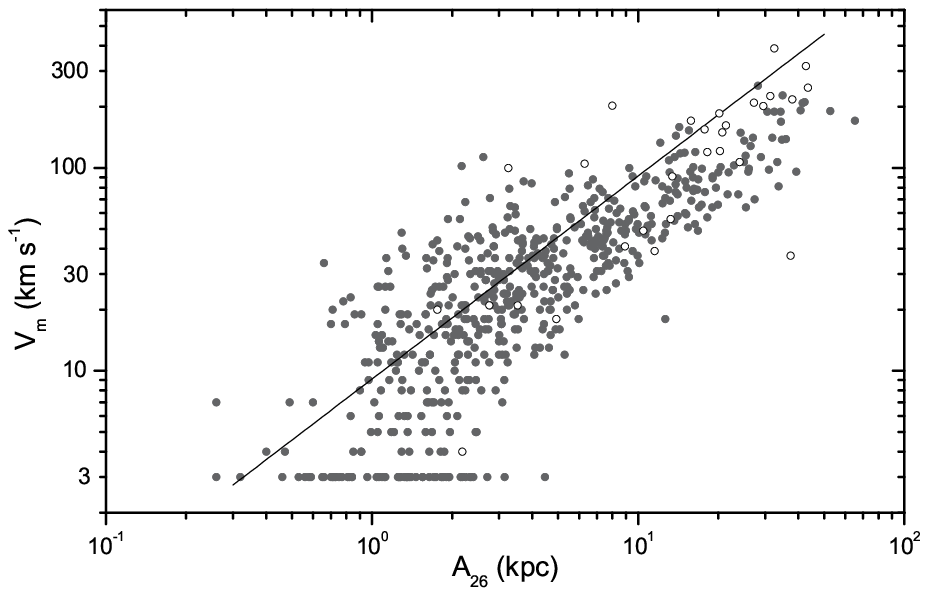} \\[-12mm]
\end{tabular}
}
\caption{Scaling relation between absolute magnitude, Holmberg diameter
	and rotation velocity for the LV galaxies. Late-type ($T > 3$)
	galaxies are shown by filled circles and early-type by open ones.}
\label{Fig9}
\end{figure*}

It is well known that both the galaxies with a dominant disk
($T>3)$, and a dominant bulge ($T<4)$ tend to be located 
along some ``fundamental plane'' in the 
parameter space \{luminosity--dimension--internal motion amplitude\}.
Three projections of this
distribution in the coordinates \{$M_B, A_{26},V_m$\}  are shown
in the panels of Fig.~\ref{Fig9}. Late-type galaxies ($T>3$) are marked by
solid circles, while the objects of early types --- with open
ones. The most distinct correlation is visible between the
linear diameter and absolute magnitude of galaxies (the bottom
panel). The straight line there corresponds to the constant
spatial luminosity density within the  Holmberg isophote,  $\log
A_{26}\propto -(2/15)M_B$, which is followed by the galaxies only
in the first approximation. Recall that diameters of faint dwarf
spheroidal satellites of the Milky Way and Andromeda were
estimated at the level, which is much fainter than the
Holmberg isophote.

Relation between the rotation amplitude of galaxy and its
absolute magnitude is shown on the middle panel. The straight
line there represents a cubic relation $L_B\sim V_m^3$, which is
also known as the blue Tully-Fisher relation, $M_B\sim-7.5\log
V_m$.  Dwarf galaxies in the lower right corner of the diagram
systematically deviate from the regression line for normal
spirals. 
In dwarf galaxies having a rotation velocity less than 30 km s$^{-1}$ the
         turbulent gas motions with typical value about 10 km s$^{-1}$ play
         significant role in mass determination. Careful account for that
         effect could straighten the Tully-Fisher relation in broader range of
         luminosity.

The upper panel of Fig.~\ref{Fig9} shows that the correlation between the
galaxy dimension and amplitude of its rotation follows the
expected linear relation  $V_m\propto A_{26}$ (direct line)
only in a rough approximation. The dispersion in this diagram is
greater than in the previous two. Perhaps one can find other
observational parameters, say, a half-luminosity radius and
a quadratic sum of rotational and turbulent velocities, which yield
this relation more linear and less scattered.

\begin{figure*}

\centerline{
\begin{tabular}{c}
\includegraphics[width=0.8\textwidth,clip]{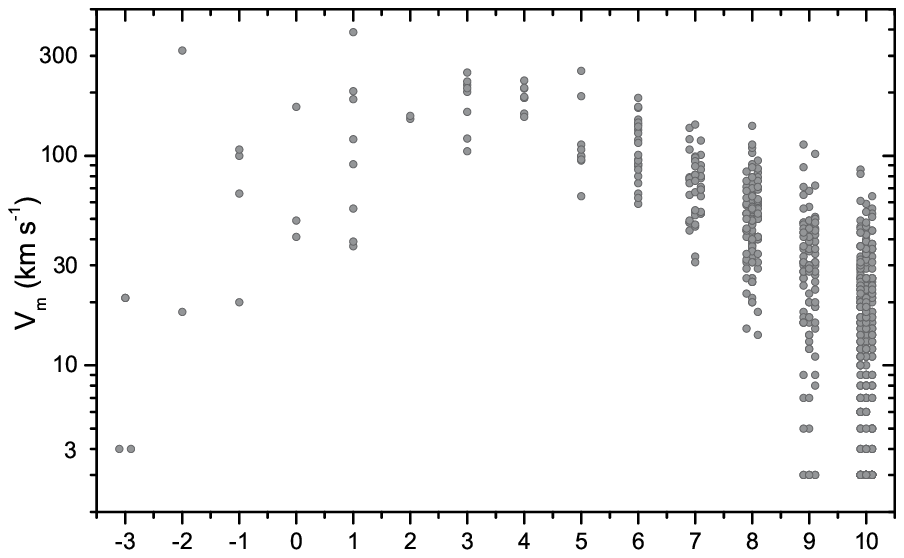} \\[-23mm]
\includegraphics[width=0.8\textwidth,clip]{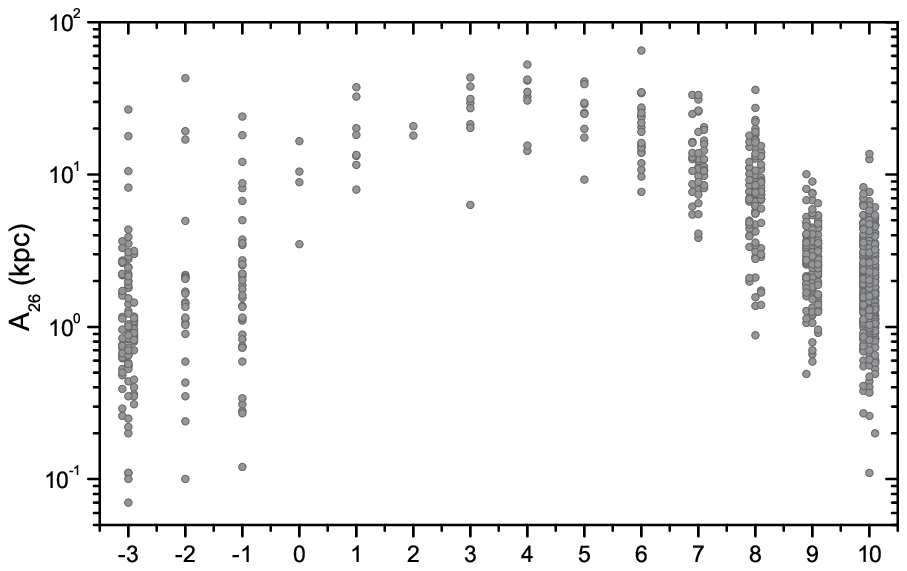} \\[-23mm]
\includegraphics[width=0.8\textwidth,clip]{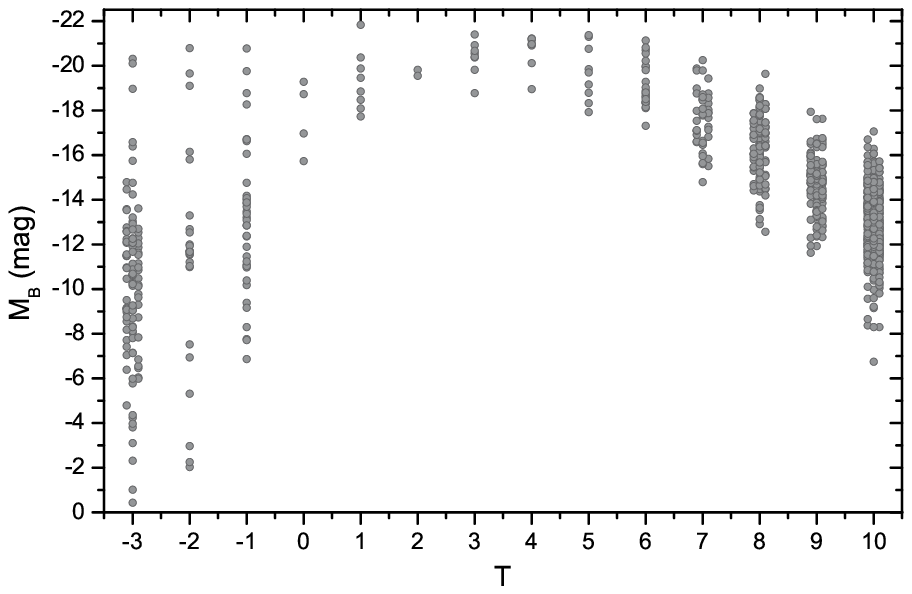} \\[-12mm]
\end{tabular}
}
\caption{Absolute magnitude, Holmberg diameter and rotation velocity of
	the LV galaxies as a function of their morphological type.}
\label{Fig10}
\end{figure*}

Luminosity, dimension and rotation amplitude are different for 
galaxies of different morphological types. Their distributions along
the $T$- scale are shown in three panels of Fig.~\ref{Fig10}. For
each global parameter: $M_B, A_{26}$ and $V_m$, the 
average value as a function of $T$ has an approximately parabolic shape
with the maximum located at type $T\simeq4$, i.e. Sbc. The scatter
of all three parameters near the maximum is minimal, which is
probably caused by specific dynamic conditions intrinsic in this type of
galaxies. Note that in the $V_m$ vs $T$ diagram, the number of
galaxies left of the peak is much smaller than at the two bottom
panels, i.e. $A_{26}$ and $M_B$. A low content of gas in 
early-type galaxies does not usually allow to measure accurate
rotation amplitude in them, which can introduce a
significant hidden selection in the analysis of diagrams similar
to Fig.~\ref{Fig9}.

\begin{figure*}

\centerline{\includegraphics[width=1.2\textwidth,clip]{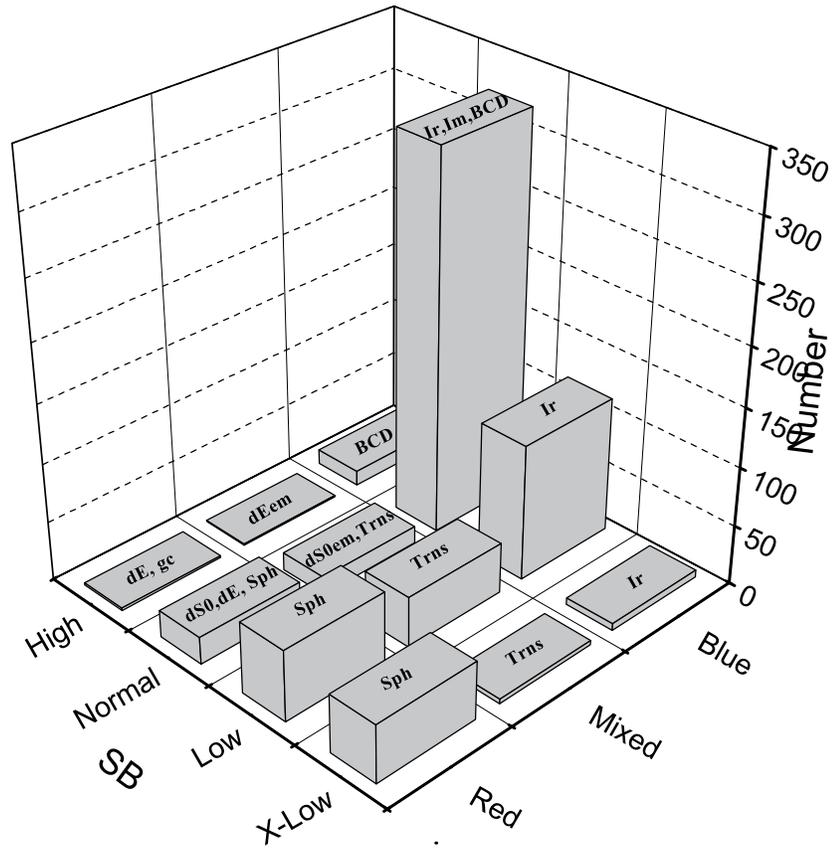}}
\caption{The same classification of dwarf galaxies, as in Fig. 2, but with
	indication of the LV galaxy number in each morphological cell.}
\label{Fig11}
\end{figure*}

\begin{figure*}

\centerline{
\begin{tabular}{c}
\includegraphics[width=0.8\textwidth,clip]{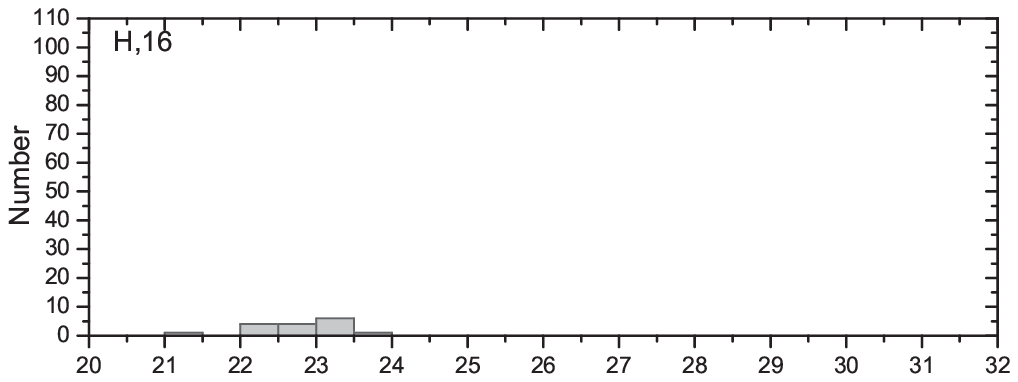} \\[-22mm]
\includegraphics[width=0.8\textwidth,clip]{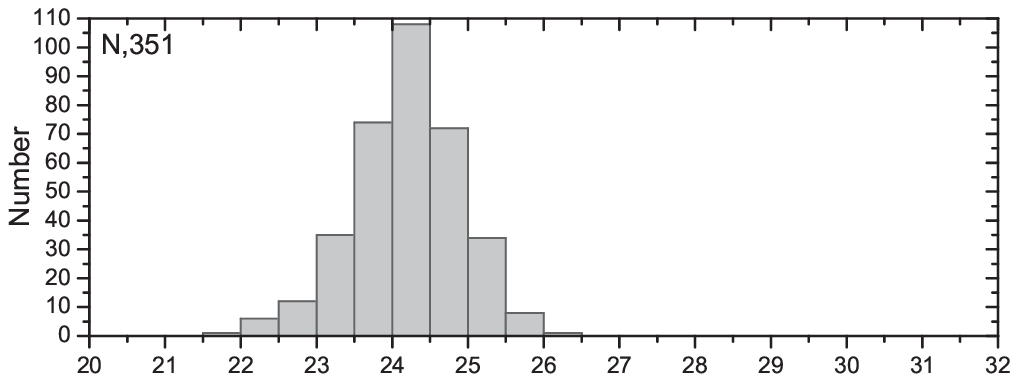} \\[-22mm]
\includegraphics[width=0.8\textwidth,clip]{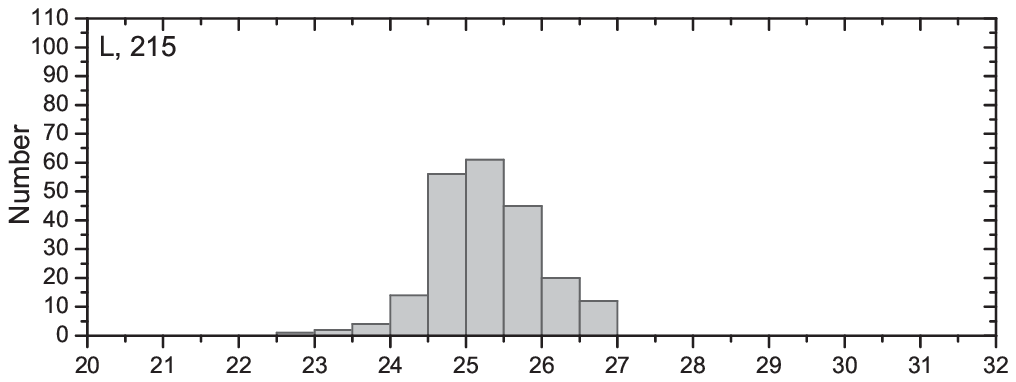} \\[-22mm]
\includegraphics[width=0.8\textwidth,clip]{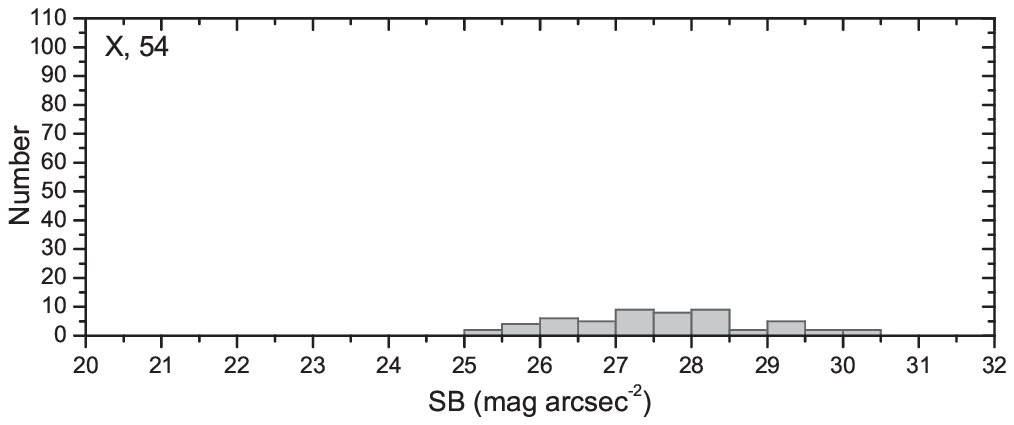} \\[-10mm]
\end{tabular}
}
\caption{Distribution of 605 dwarf galaxies in the LV according to their
	mean surface brightness. Galaxies classified as having extremely
	low (X), low (L), normal (N) and high (H) surface brightness are
	presented on different panels.}
\label{Fig12}
\end{figure*}

It should be noted that in the Hubble morphological classification
scheme, the type of dwarf irregular galaxies was a sort of a
``trash bin'' where the ``ugly'' objects were dropped. However,
3/4 of the sample of the LV consist of dwarf galaxies.
One can see from the previous figure that the distribution of 
$M_B, A_{26}$ and $V_m$ parameters for them merges into a thick
lane, denoting the need of a more refined classification of dwarf
systems. The two-dimensional classification (Fig.~\ref{Fig2}) we propose
is based on the difference in their color (or the presence of
emission) and surface brightness. Fig.~\ref{Fig11} shows
distribution of number of dwarf objects by type in our
scheme. The cells of the scheme vary considerably in the degree
of filling, which is obviously caused by conditions of
formation  of dwarf systems and their subsequent evolution in
different environments. However, the 2D-shape of this diagram may
be subject to a significant selection effect: the galaxies of
extremely low surface brightness and poor in gas, such as
Apples~I, can easily stay undetected both in the optical and HI
surveys. It follows from the histograms of Fig.~\ref{Fig12} that though
the distributions of dwarf galaxies by the visual surface
brightness gradations overlap, they have a distinct median trend
on the SB scale [mag arcsec$^{-2}$]: 23.0 (H), 24.2 (N), 25.2 (L)
and 27.6 (X). Observed differences in the mean surface brightness of
dwarf galaxies reach more than 10 magnitudes, and these
differences are obviously caused by features of star
formation history in them.

\begin{figure*}

\centerline{
\begin{tabular}{c}
\includegraphics[width=0.8\textwidth,clip]{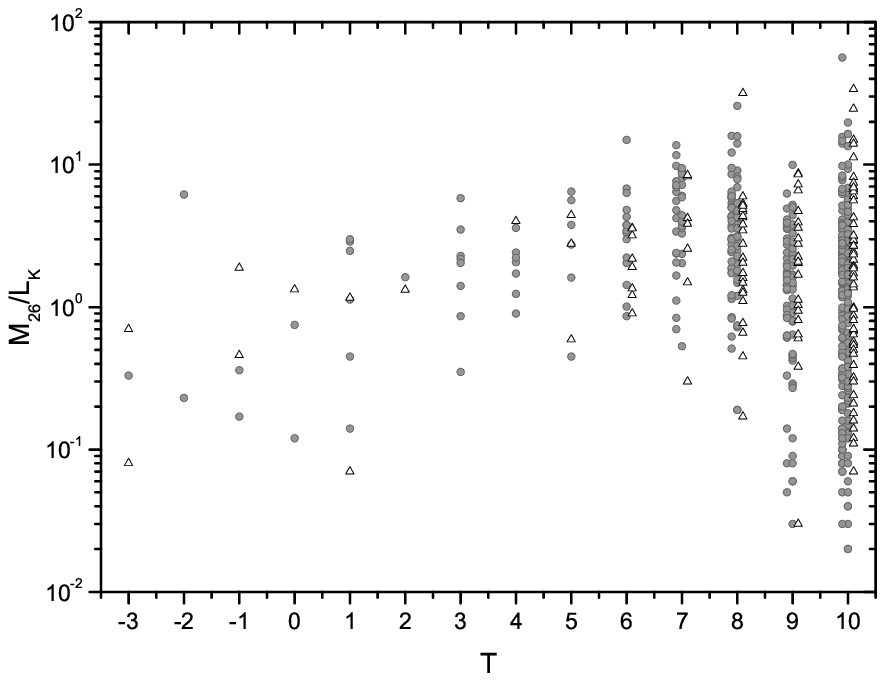} \\[-22mm]
\includegraphics[width=0.8\textwidth,clip]{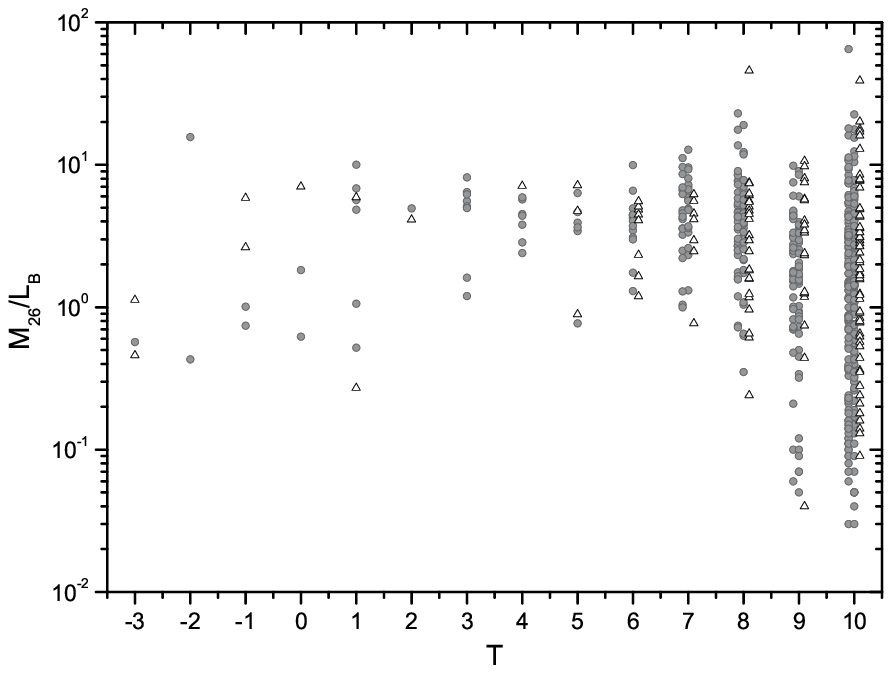} \\[-10mm]
\end{tabular}
}
\caption{Indicative mass-to-luminosity ratio in B- and K-bands as a function
	of morphological type of the LV galaxies. Face-on galaxies with
	inclination  $i < 45\degr$ are shown by open triangles.}
\label{Fig13}
\end{figure*}

An important dynamic characteristic of galaxies is the ratio of
their indicative mass within the Holmberg isophote to the total
luminosity. Fig.~\ref{Fig13} represents distribution of this ratio in
the B- and K-bands for different morphological types. The galaxies
with inclination angles $i<45^{\circ}$, for which the
estimate of $M_{26}$ may contain a considerable uncertainty, are
marked with empty symbols. Both ratios show a growth trend from
early to late types, more pronounced in the case of ratio
 $M_{26}/L_K$. However, in the transition from spiral galaxies
(Sd, Sm) to irregulars (BCD, Im, Ir), there is a significant (about 4 times)
decrease in the average $M_{26}/L$ ratio. This effect may be due
to a difference in structure and kinematics of galaxies in
the presence or absence of spiral pattern, in
particular owing to the contribution of turbulent motions,
ignored in the evaluation of mass $M_{26}$ by equation (9).

\section{Main HI properties of the LV sample}

Hydrogen characteristics of different samples of galaxies,
limited by HI flux, apparent magnitude or angular diameter of
galaxies have repeatedly been investigated (Huchtmeier \& Richter
1988, Roberts \& Haynes 1994, Zwaan et al., 2003, Martin et al. ,
2010). These samples have a strong bias towards the disk-shaped
galaxies of high luminosity and do not reflect the HI-properties
of galaxies in the unit volume, dominated by dwarf objects.
Currently, more than 70\% of the LV galaxies are
detected in the HI line, and an upper limit of HI fluxes is
known for another $\sim14$\% of these galaxies. A huge progress
has been made here via the massive HI surveys on the Parks
(HIPASS) and Arecibo (ALFALFA) radio telescopes, and within a
special survey of nearby dwarf galaxies, performed by Huchtmeier
in Effelsberg. However, at high declinations (Dec$>+38^{\circ}$)
there is still a number of nearby galaxies, not observed in the
HI line. The planned ``blind'' HI-survey of the northern sky
(WNSHS) in Westerbork (http://www.astron.nl/~jozsa/wnshs/) will
obviously fill this gap shortly.

\begin{figure*}

\centerline{
\begin{tabular}{c}
\includegraphics[width=0.8\textwidth,clip]{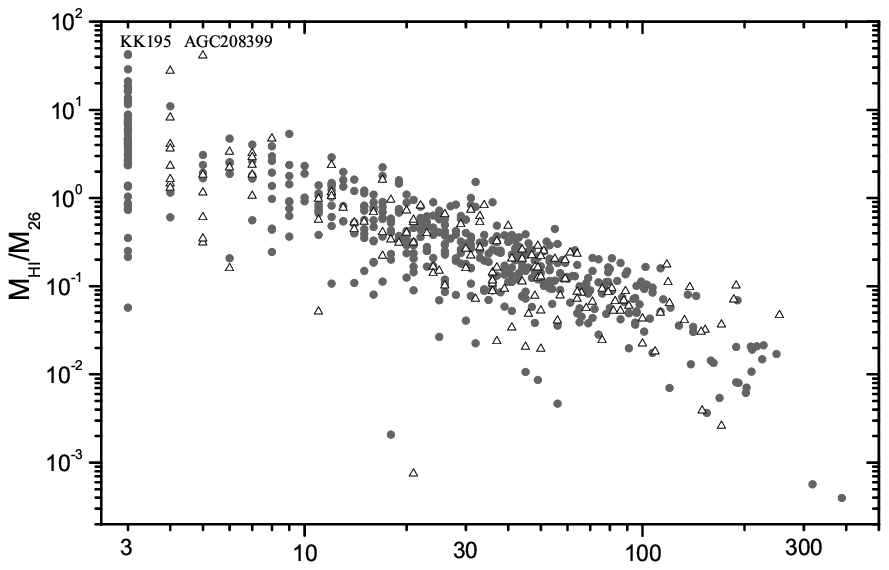} \\[-22mm]
\includegraphics[width=0.8\textwidth,clip]{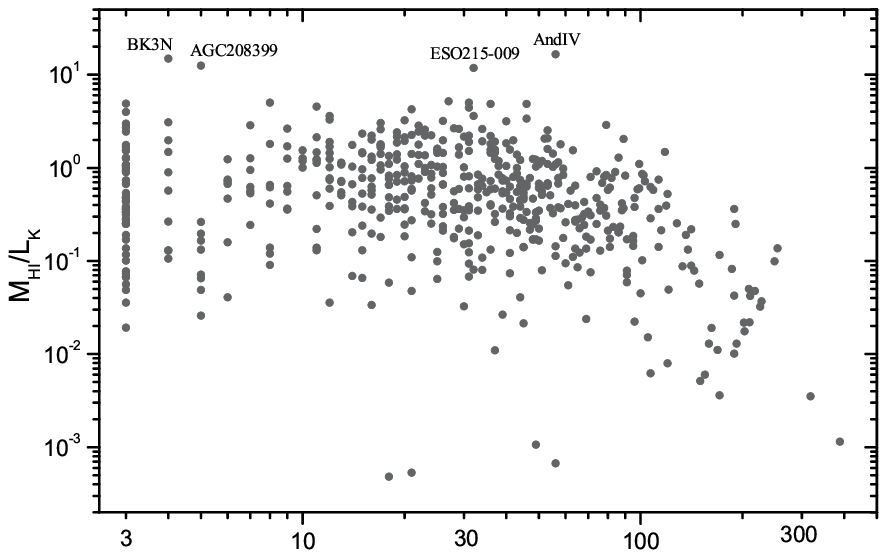} \\[-22mm]
\includegraphics[width=0.8\textwidth,clip]{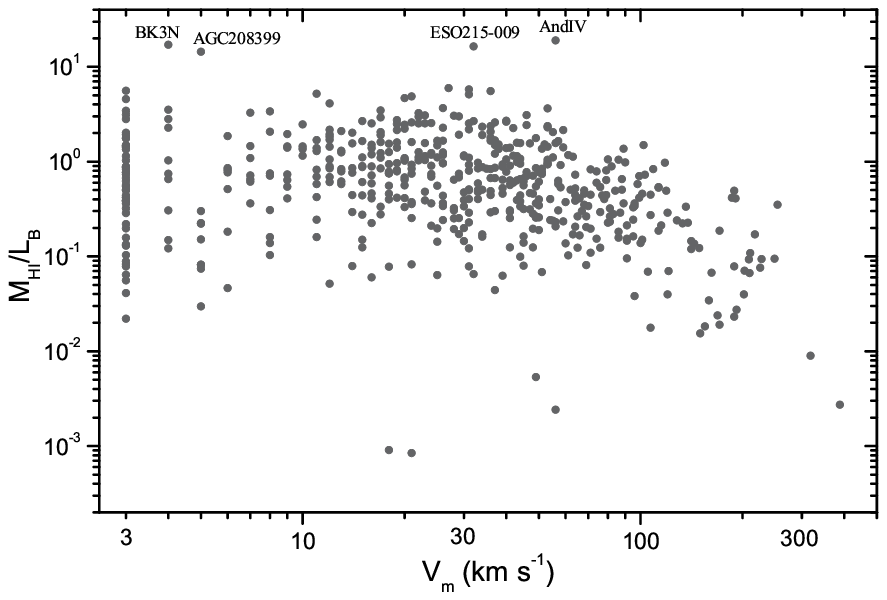} \\[-10mm]
\end{tabular}
}
\caption{Hydrogen mass-to-luminosity ratio in B- (bottom) and K- band
	(middle) versus rotation velocity of the LV galaxies. The upper
	panel presents hydrogen mass-to-indicative mass for them. Face-on
	galaxies are marked by open triangles.}
\label{Fig14}
\end{figure*}

Fig.~\ref{Fig14} represents the ratio of hydrogen mass to
B- and K-band luminosity versus the rotation amplitude
of galaxies $V_m$ (lower and middle panel). The
sample of nearby galaxies demonstrates the well-known effect that
the average ratio  $M_{HI}/L_B$ increases from
$\sim0.1M_{\odot}/L_{\odot}$ to $\sim0.7M_{\odot}/L_{\odot}$
in the transition from normal spiral galaxies to dwarf systems
with $V_m<50$ km s$^{-1}$. This feature is usually attributed
to the lower rate of star formation in dwarf galaxies with their
turbulent motions compared to the disks, where regular rotation
and density waves boost the process of star formation. Note that
the smallest dwarfs with $V_m<6$ km s$^{-1}$ show some tendency of
decreasing the $M_{HI}/L$ ratio. Perhaps this is due to the fact
that a shallow potential well in dwarf systems is not capable of
retaining large masses of gas.

The K-band luminosity of galaxies is a good indicator of their
stellar mass, $M_*$, since it is almost insensitive to 
internal extinction and the presence of young stellar population.
At  $M_*/L_K\simeq 1M_{\odot}/L_{\odot}$ (Bell et al. 2001) the
$M_{HI}/L_K$ ratio in the middle panel of Fig.~\ref{Fig14} actually is
equivalent to the mass ratio of gas and stars in the galaxies. For
dwarf galaxies with $V_m<50$ km s$^{-1}$,  the average
$M_{HI}/L_K$ ratio amounts to $\sim0.7M_{\odot}/L_{\odot}$.
Taking into account a correction for abundances of helium
and molecular gas, $M_{gas}=1.85M_{HI}$ (Fukugita \& Peebles
2001), we obtain the characteristic ratio of $M_{gas}/M_*=1.3$
for dwarf galaxies. Hence, more than a half of the baryon mass in
dwarf galaxies, detected in HI remains unprocessed into the
stellar component.

The upper panel of Fig.~\ref{Fig14} shows behavior of the hydrogen
mass-to-dynamical mass ratio within a Holmberg radius depending
on $V_m$. For normal spiral galaxies ($V_m>100$ km s$^{-1}$) we
have the median value of $M_{HI}/M_{26}\sim0.03$, while for the
dwarf galaxies with $V_m<20$ km s$^{-1}$ it increases to
$M_{HI}/M_{26}\sim1$, in some cases reaching   $M_{HI}/M_{26}>10$.
Only a small fraction of these extreme ratios is due to 
errors in determining the inclination angle of galaxy  $i$
(the cases with  $i<45^{\circ}$ are marked in this panel with
empty triangles).

\begin{figure*}

\centerline{
\begin{tabular}{c}
\includegraphics[width=0.8\textwidth,clip]{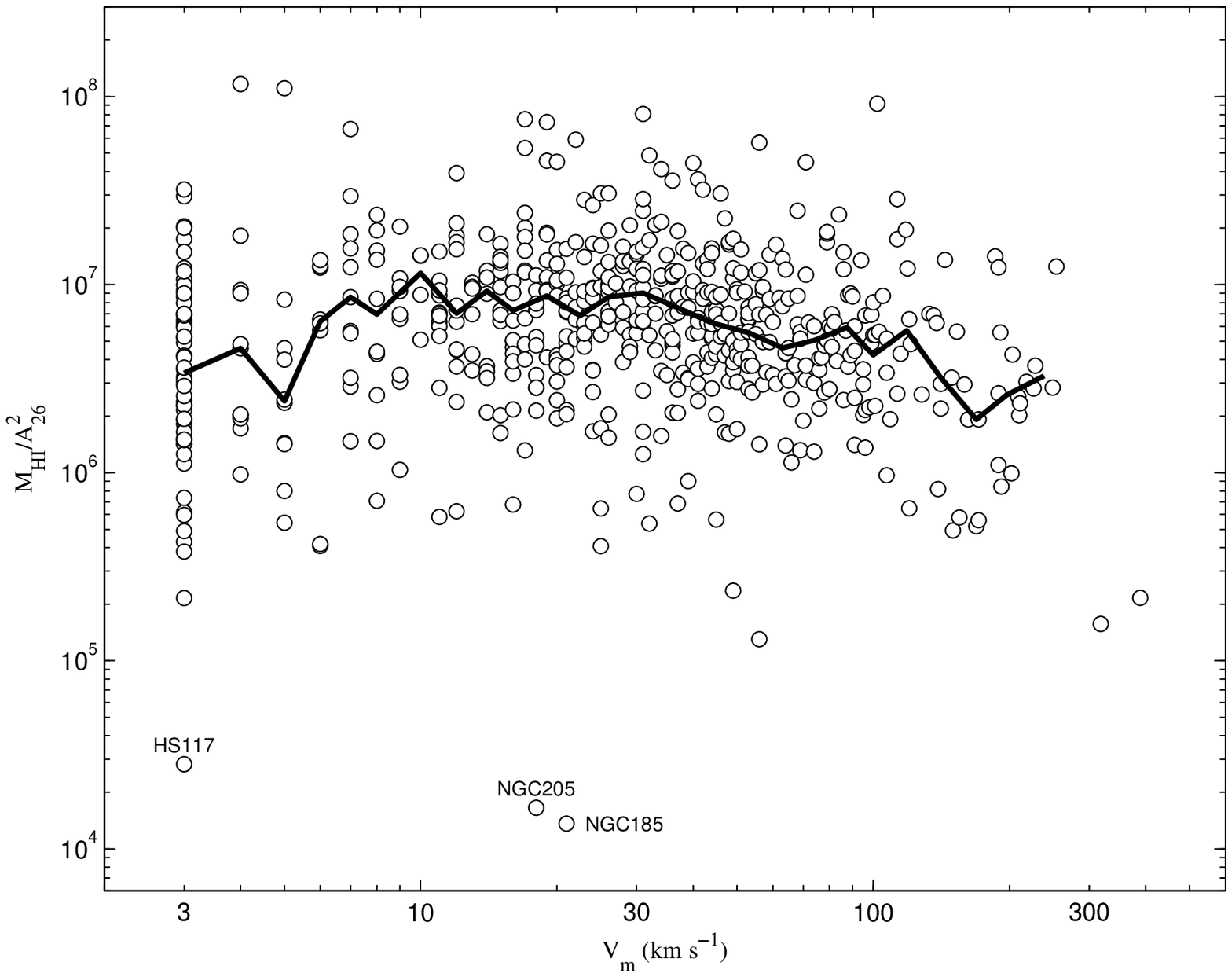} \\[-12mm]
\includegraphics[width=1.0\textwidth,clip]{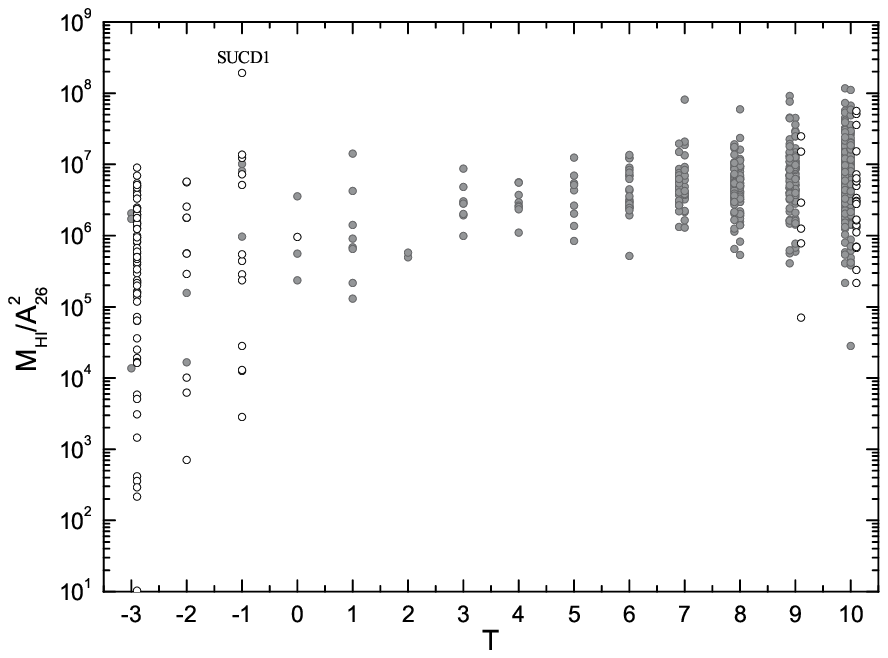} \\[-10mm]
\end{tabular}
}
\caption{
Relation between mean surface density of hydrogen mass and rotation velocity for LV galaxies
is shown on the top panel. The thick line represents the running median.
Dependence of mean surface density versus morphological type is presented on bottom panel.
Objects	with the upper limit of hydrogen mass are shown by open circles.
}
\label{Fig15}
\end{figure*}

 As noted by Roberts \& Haynes (1994), an important global parameter,
characterizing the conditions of star formation in a galaxy is its
average surface density of hydrogen, $M_{HI}/A_{26}^2$. The
distribution of this parameter for the LV galaxies is
represented in the panels of Fig.~\ref{Fig15} depending on the rotation amplitude
$V_m$ and morphological type $T$. In the $T\geq0$
region, the mean density   $M_{HI}/A_{26}^2$ weakly increases
towards the late types, showing the minimum variance around $T=4$
(Sbc). Large variations of surface density of hydrogen mass
at both edges of the morphological scale is easily explained by
some features of their evolution: the depletion of gas
reserves in E, S0-galaxies, the sweep-out of gas from 
dwarf systems during bursts of star formation and/or under their
passage through the halo of massive galaxies. For the faintest
dwarf galaxies with rotation velocities of $V_m<6$ km s$^{-1}$,
some noticeable decline of the average surface density of hydrogen
mass is observed.

\section{Environment and gas-to-star transformation}

Different indicators can be used for quantitative description
of a galaxy environment density. Karachentsev \& Makarov
(1999), considering a galaxy ``i'', performed ranking of its
neighbors ``n'' by a tidal force magnitude $F_n\sim
M_n/D^3_{in}$, where  $D_{in}$ is a spatial distance of 
neighboring galaxy, and $M_n$ is its mass, which is considered
proportional to the galaxy luminosity. The local mass density was
characterized by the ``tidal'' index
$$\Theta=\max[\log(M_n/D_{in}^3)]+C, \,\,\,\, n=1, 2\ldots N \eqno(12) $$
of the most significant neighbor, called the ``Main Disturber''
(= MD). The value of constant $C= -10.96$ was chosen so that at
$\Theta=0$ the galaxy ``i'' is located on the ``zero velocity
sphere'' relative to the MD. At that, the galaxies with $\Theta>0$
turned out to be members of a certain group, and the negative
values of $\Theta$ corresponded to isolated galaxies.

For each galaxy of the LV, we calculated the tidal
index $\Theta_1$ according to (12), considering the masses of
galaxies to be proportional to their K-luminosities. The value of
 $\Theta_1$ index and the name of Main Disturber are presented
in columns (13) and (14) of Table~\ref{t:param}. The tidal index $\Theta_1$
or actually a stellar density contrast, contributed by one, most
important neighbor, i.e. MD, can significantly change with time
due to orbital motions of galaxies. It therefore seems
reasonable to use other parameters for characterizing the local
environment. As a case of another, more robust
estimator, we calculated the index
$$\Theta_5=\log(\sum^5_{n=1} M_n/D_{in}^3)+C, \eqno(13)$$
which is the sum of the density contrasts produced by five most
important neighbors. The value of constant $C$ here is the
same as in (12).

Finally, in the last column of Table~\ref{t:param} we present the third
environment estimator ---  logarithm of the average density of
K-luminosity of galaxies within the 1 Mpc radius sphere around
the galaxy under consideration, expressed in units of the global
mean density of K-luminosity, $4.28\cdot10^8L_{\odot}/{\rm Mpc}^3$, derived from 2MASS
(Jones et al. 2006):
$$ \Theta_j=\log(j_K(1{\rm Mpc})/j_{K,global}). \eqno(14) $$
Here, the central galaxy is not included in the estimate of
average density. Note that in some galaxies no neighbors were
detected within 1 Mpc and the $\Theta_j$ index for them is
formally assumed to be $-3.0$.

\begin{figure*}

\centerline{
\begin{tabular}{c}
\includegraphics[width=0.8\textwidth,clip]{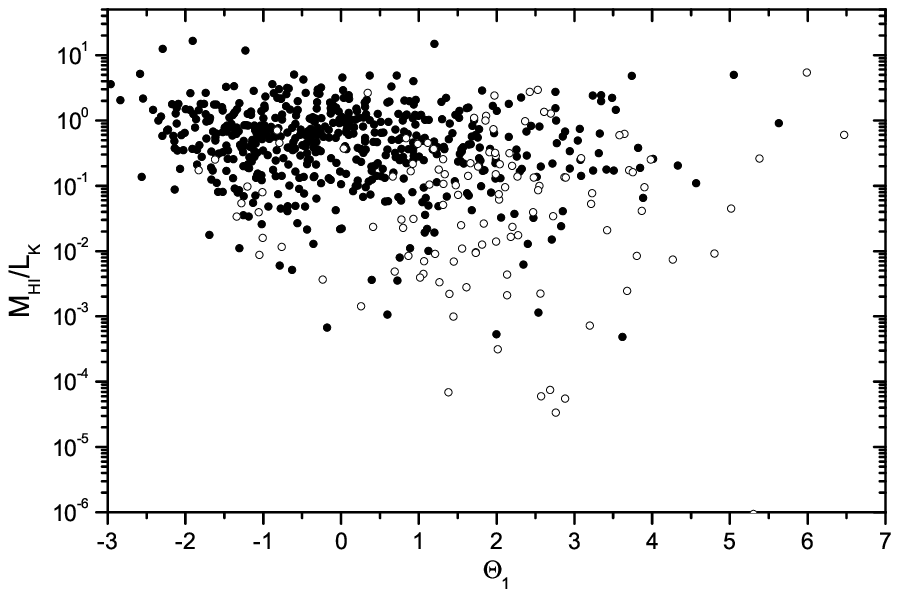} \\[-24mm]
\includegraphics[width=0.8\textwidth,clip]{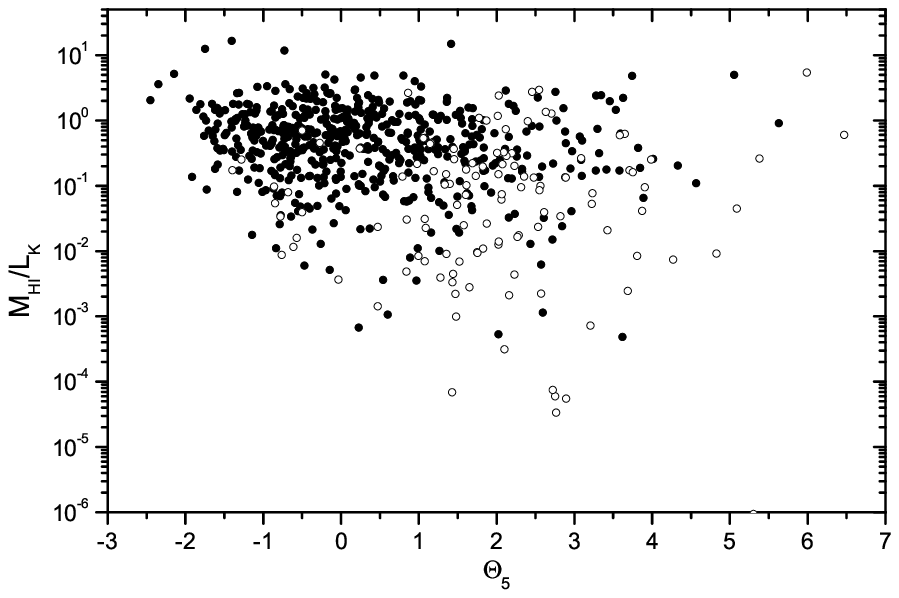} \\[-24mm]
\includegraphics[width=0.8\textwidth,clip]{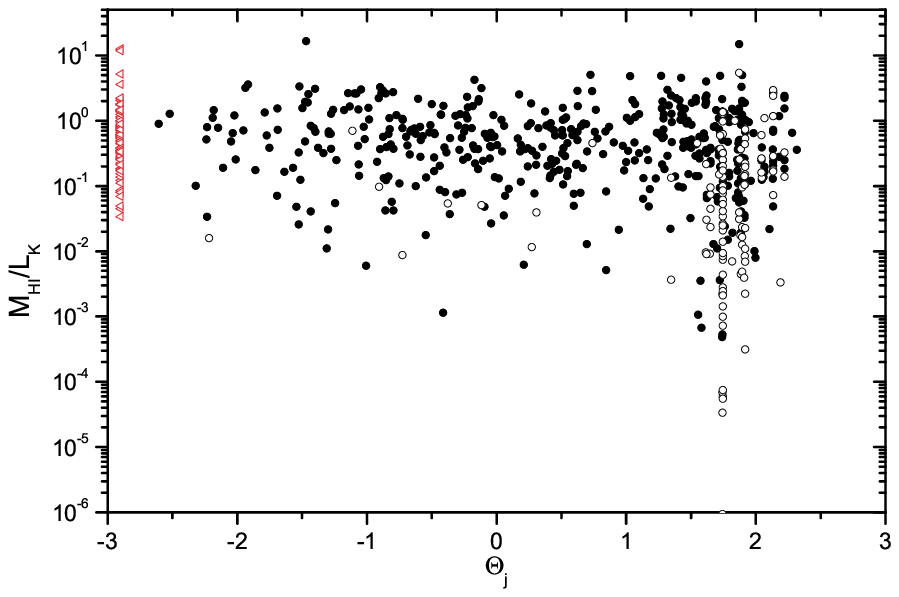} \\[-12mm]
\end{tabular}
}
\caption{Hydrogen mass-to-K-band-luminosity ratio vs. different estimators
	of the local environment density, i.e. tidal index. Galaxies with
	a positive tidal index correspond usually to group members. Open
	circles indicate galaxies with only upper limit of the HI flux.}
\label{Fig16}
\end{figure*}

Distributions of galaxies in the LV by hydrogen
mass-to-K-luminosity ratio and tidal index $\Theta_1,
\Theta_5$, or $\Theta_j$  are presented in Fig.~\ref{Fig16}. Galaxies
with the HI flux upper limit are shown by open circles. These
data show that the maximum value of $M_{HI}/L_K$ 
remains approximately the same both for the field galaxies and
group members, regardless of the adopted density estimator:
$\Theta_1, \Theta_5$ or $\Theta_j $. In contrast, there is a
steep drop in the minimum value of $M_{HI}/L_K$ in the direction
of high-density environments. It is obvious that the observed
increase in HI deficiency towards high-density
environment is due to the sweep-out of gas from galaxies under
their tight interaction. This process can reduce
the amount of hydrogen in a galaxy by 2--4 orders, thereby
affecting the rate of star formation in it.

To estimate the integral star formation rate in a galaxy, we
used the relation
$$ SFR[M_{\odot} yr^{-1}]=0.945\cdot10^9F_c(H\alpha)D^2 \eqno(15)$$
from Kennicutt (1998), where $D$ is the distance in Mpc, and
$F_c(H\alpha)$ is the integral flux in the $H\alpha$ line [erg
cm$^2$s$^{-1}$],   corrected for Galactic and internal
extinction as
$$A(H\alpha)=0.538(A^G_B+A^i_B). \eqno(16)$$
Values of the  $F(H\alpha)$ flux and apparent 
$m(H\alpha)$ magnitudes in Table~\ref{t:lv} are related by (3).

\begin{figure*}

\centerline{
\begin{tabular}{c}
\includegraphics[width=0.8\textwidth,clip]{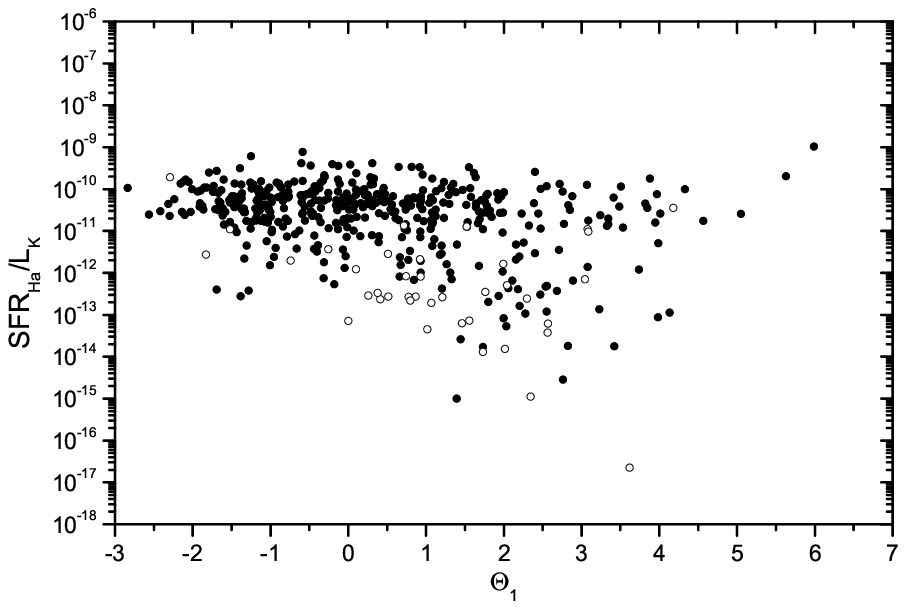} \\[-22mm]
\includegraphics[width=0.8\textwidth,clip]{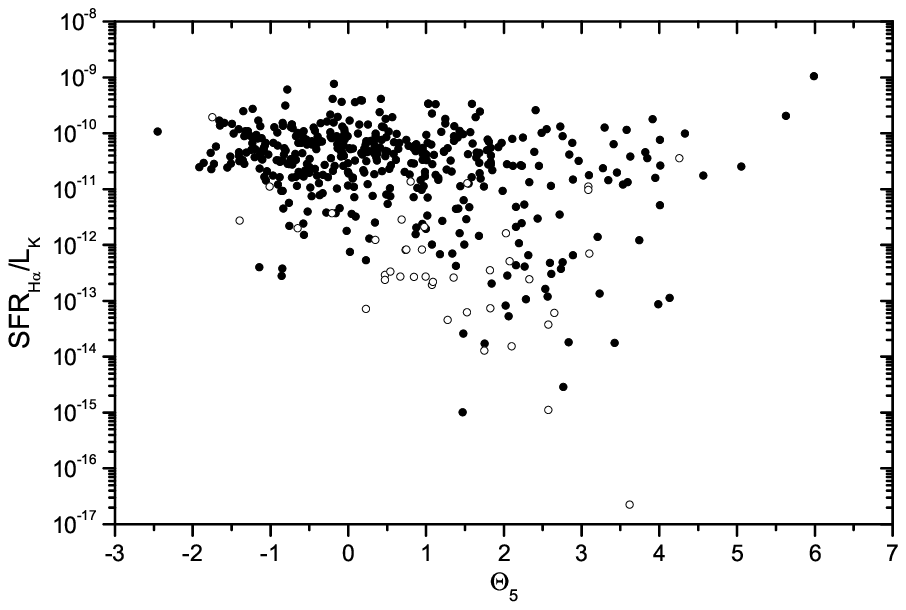} \\[-22mm]
\includegraphics[width=0.8\textwidth,clip]{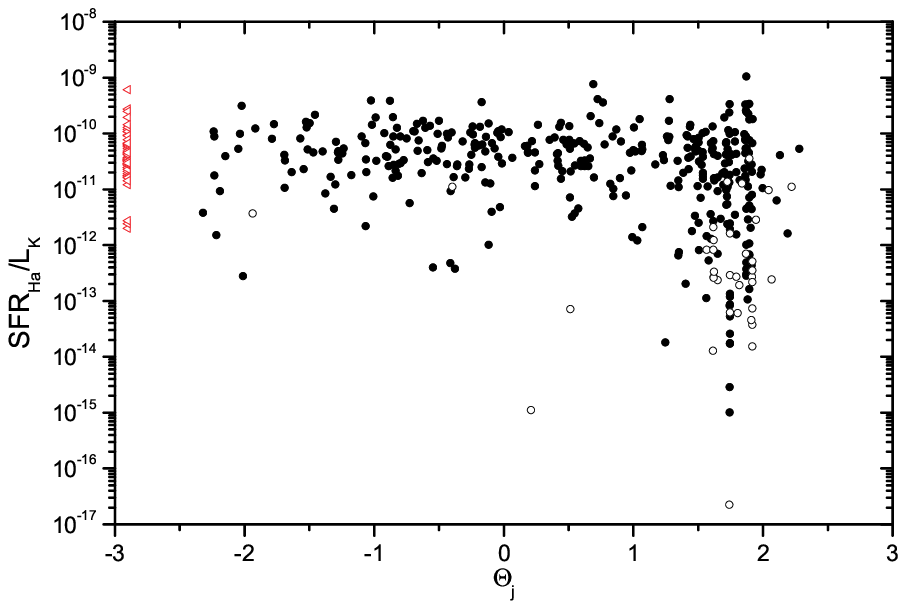} \\[-12mm]
\end{tabular}
}
\caption{Specific star formation rate for 493 LV galaxies determined
	from their $H\alpha$ flux versus different kinds of tidal index.
	Galaxies with upper limit of $H\alpha$ flux are indicated by open
	circles.}
\label{Fig17}
\end{figure*}

Distribution of the specific star formation rate per unit of
K-luminosity of galaxy is presented in three panels of
Fig.~\ref{Fig17}, depending on $\Theta_1, \Theta_5$ and $\Theta_j$ as an
argument. The shape of these diagrams is largely similar to the
corresponding  $M_{HI}/L_K$ distributions  in Fig.~\ref{Fig16}. The
increasing scatter of both  $M_{HI}/L_K$, and $SFR/L_K$
from isolated galaxies to group members gives evidence for a
well known fact that the environment of galaxies significantly
affects the process of converting gas into stars.

To date, $H\alpha$ fluxes are measured for  495
galaxies in the LV. The SFR estimates via
$H\alpha$ correspond to a time scale of $\sim10$ Myr, characteristic
for the glow of the most massive stars. Another
SFR estimate  can be obtained from the integral FUV-flux of
galaxy, the values of which in the form of $m_{FUV}$-magnitudes are
presented in Table~\ref{t:lv}. Following Lee et al. (2011), we have used
the relation
$$\log(SFR[M_{\odot}\cdot{\rm yr}^{-1}])=2.78-0.4m^c_{FUV}+2\log D, \eqno(17)$$
where $D$ is in Mpc, and the apparent FUV-magnitude is
corrected for extinction as
$$m^c_{FUV}=m_{FUV}-1.93(A^G_B+A^i_B). \eqno(18)$$

\begin{figure*}

\centerline{
\begin{tabular}{c}
\includegraphics[width=0.8\textwidth,clip]{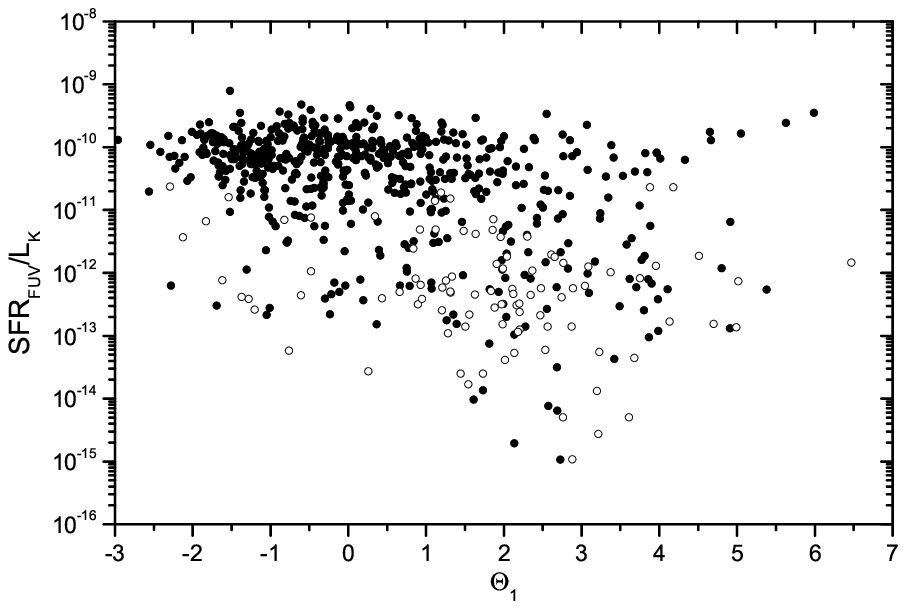} \\[-22mm]
\includegraphics[width=0.8\textwidth,clip]{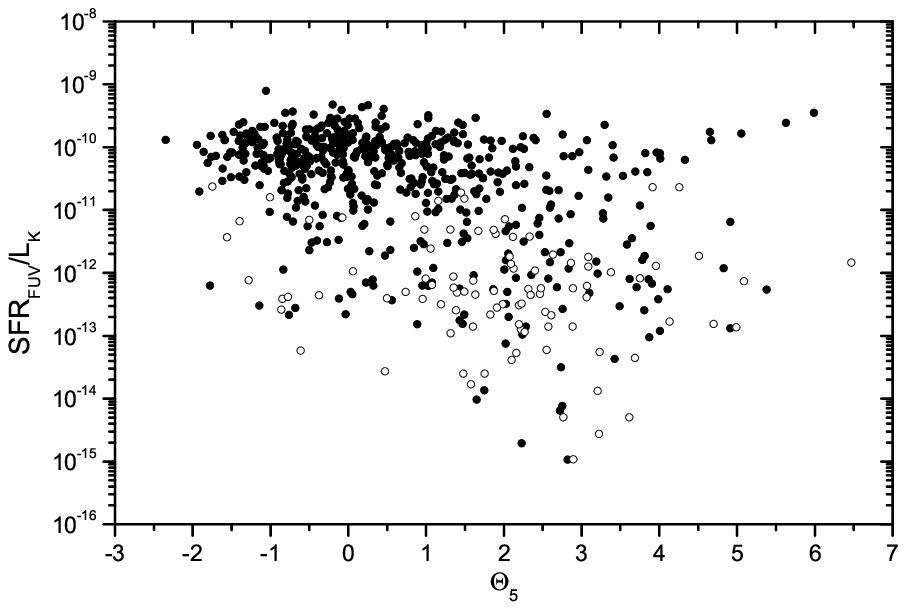} \\[-22mm]
\includegraphics[width=0.8\textwidth,clip]{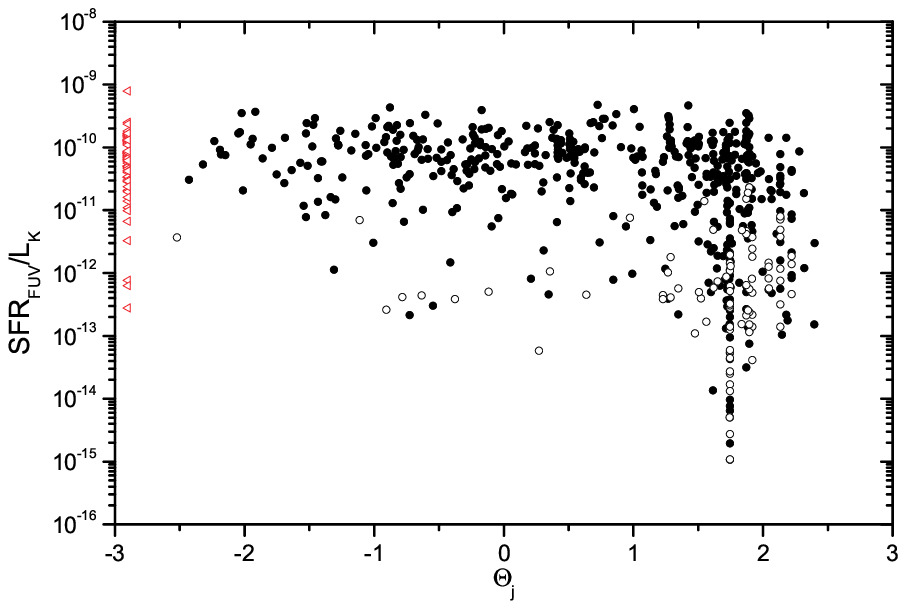} \\[-10mm]
\end{tabular}
}
\caption{Specific star formation rate for 692 LV galaxies derived from
	their FUV-flux versus three kinds of tidal index. Open circles
	indicate galaxies with upper limit of FUV-flux.}
\label{Fig18}
\end{figure*}

The SFR values  from (17) and (18) refer to a characteristic glow
time $\sim100$ Myr, i.e. they are more robust. However,
they are much more subject to uncertainty due to poorly known
internal extinction in galaxies. Distributions of the
specific  star formation rate for 715 galaxies on the scales
$\Theta_1, \Theta_5$ and $\Theta_j$  (Fig.~\ref{Fig18}) reproduce in general
the previous distributions. Here, alike the previous
figure, the galaxies with upper limit of SFR are marked by
open circles.

\begin{figure*}

\centerline{
\begin{tabular}{c}
\includegraphics[width=0.8\textwidth,clip]{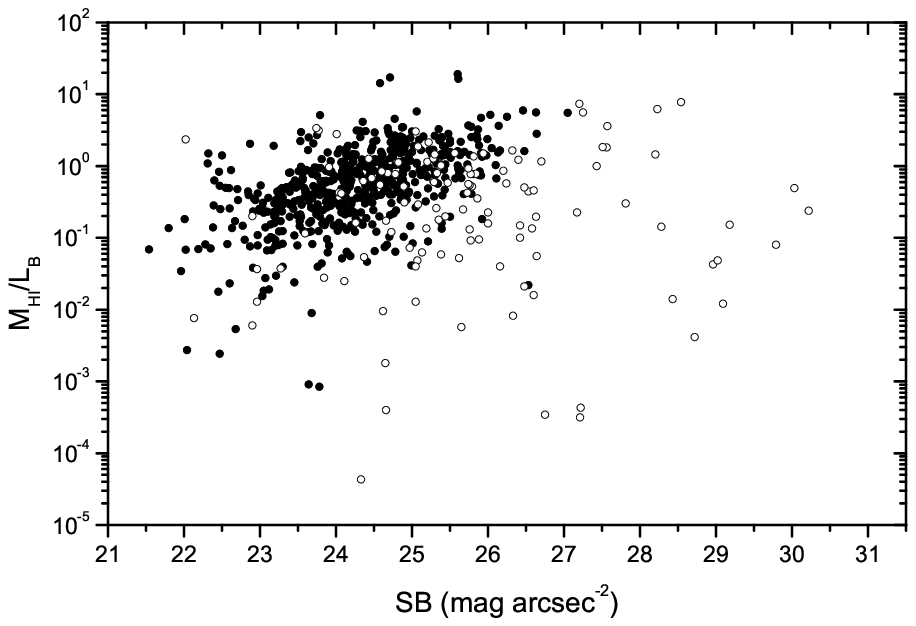} \\[-23mm]
\includegraphics[width=0.8\textwidth,clip]{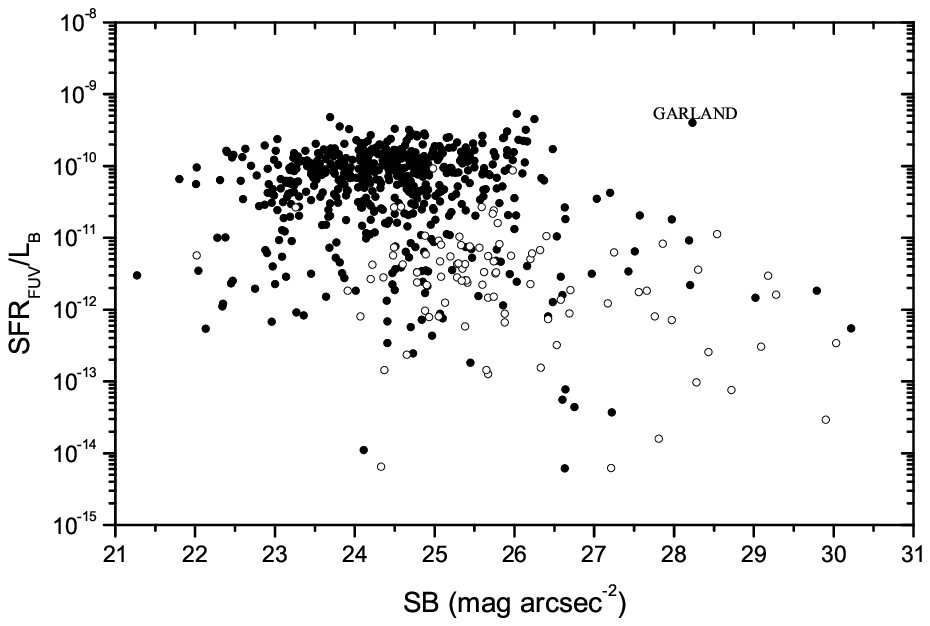} \\[-23mm]
\includegraphics[width=0.8\textwidth,clip]{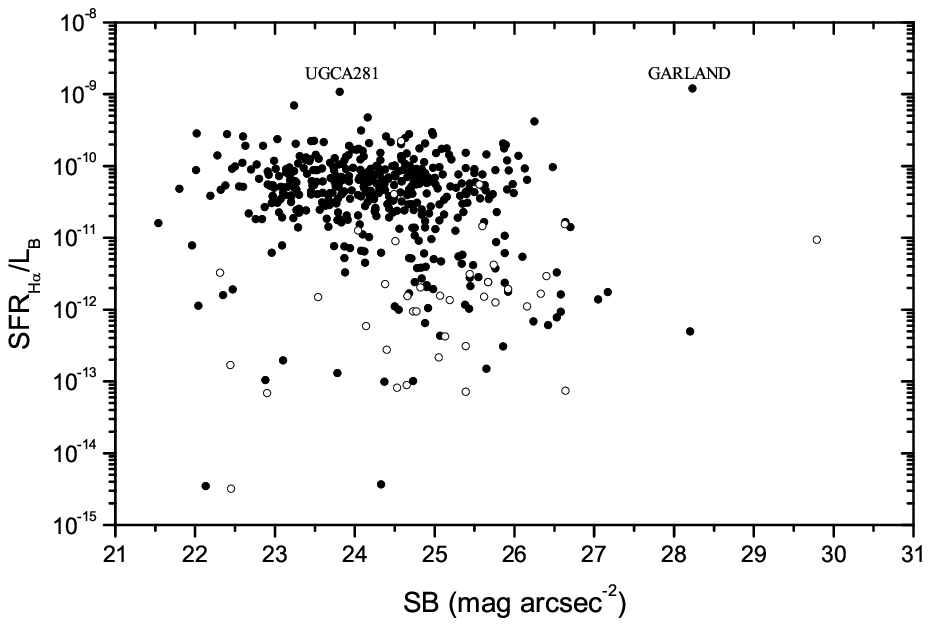} \\[-12mm]
\end{tabular}
}
\caption{Specific star formation rate and hydrogen mass-to-luminosity
	ration in the B-band versus mean surface brightnes of the LV
	galaxies. Open circles indicate galaxies with the upper limit
	of corresponding fluxes.}
\label{Fig19}
\end{figure*}

There sometimes occur certain reasonings in the literature,
stating that the galaxies of low and normal surface brightness
significantly differ in their evolutionary history. As follows
from the data of Fig.~\ref{Fig19}, the specific star formation rate,
measured by the $H\alpha$ and FUV fluxes is practically
independent on the mean surface brightness of galaxies in the
range of about five magnitudes. The SFR decline becomes noticeable
only for the extremely low surface brightness galaxies. The
relative hydrogen abundance  $M_{HI}/L_B$ grows towards
faint surface brightnesses, but at $SB>27$ the statistics of 
observables gets poor. These empirical relationships are
obviously in need of theoretical interpretation.

\section{Some local vs. global parameters}

According to Sloan Digital Sky Survey, the cosmic variation of
luminosity in the cubic cell with a 30 Mpc side amounts to
$\sim30$\% and drops to a value of $\sim10$\% in a cell with an
edge of 95 Mpc (Papai \& Szapudi 2010). The local sphere of 20
Mpc diameter contains a number of groups of different populations
and morphology, as well as some voids, almost completely devoid of galaxies.
It is therefore important to have a quantitative idea of how
drastically parameters of the LV differ from the
global cosmic parameters.

\begin{figure*}

\centerline{\includegraphics[width=0.8\textwidth,clip]{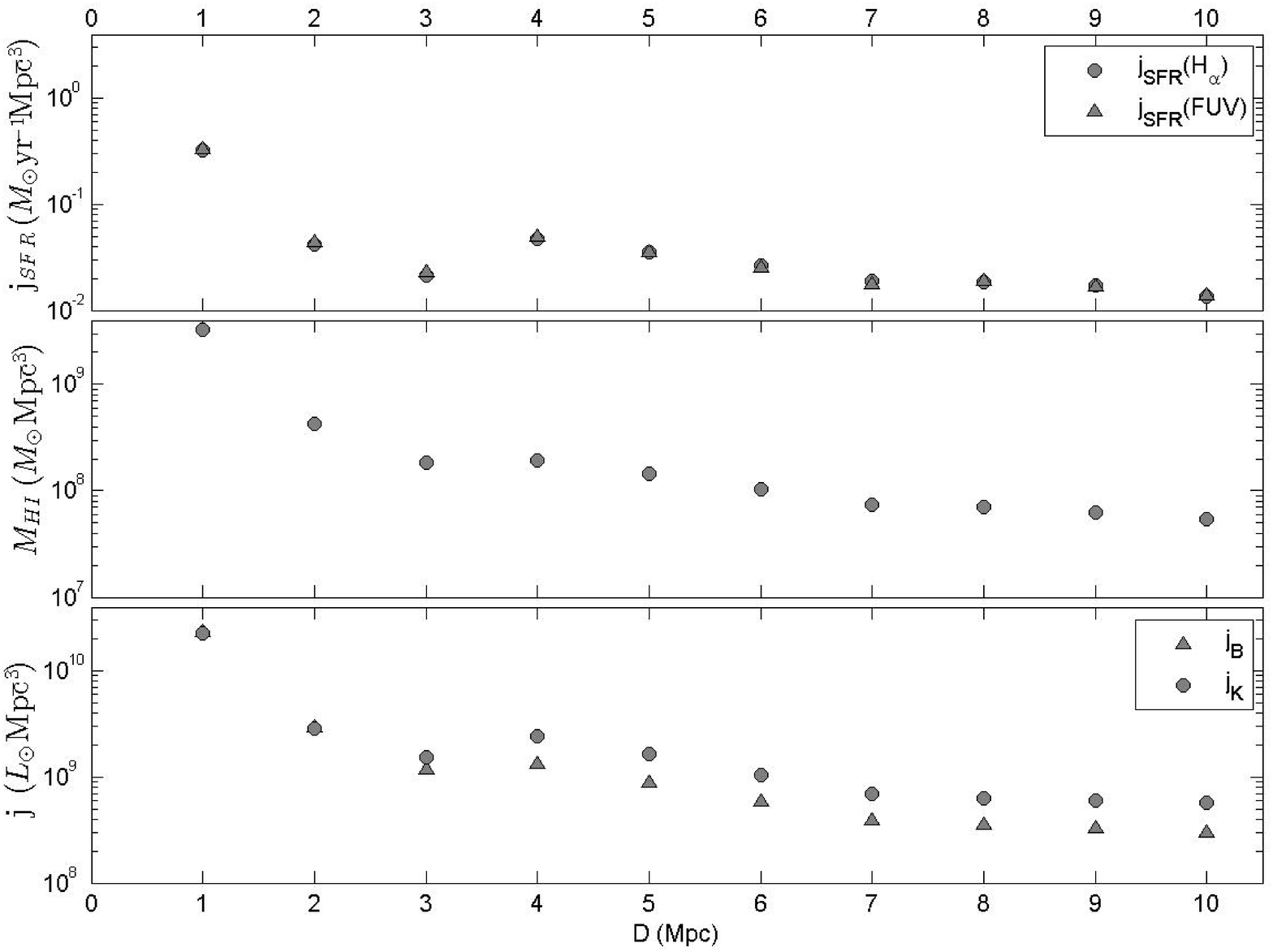}}
\caption{Average density of luminosity in B- and K- bands within a distance
	D around our Galaxy (bottom). The same for the hydrogen mass density
	(middle), and for the star formation rate (top) estimated
	either from $H\alpha$ fluxes or FUV-fluxes.}
\label{Fig20}
\end{figure*}

The bottom panel of Fig.~\ref{Fig20} shows behavior of the average
luminosity density in the B-band (triangles) and K-band (circles)
inside spheres of fixed radii. Within a radius of $D=10$ Mpc the average
B-luminosity density $3.0\cdot10^8L_{\odot}/$ Mpc$^3$ is 2.2 times
higher than its global value  $(1.3\pm0.1)\cdot10^8L_{\odot}/$
Mpc$^3$ according to Blanton et al. (2003) and Liske et al.
(2003). For the  K-luminosity density, the local value 
$j_K(D<10$ Mpc) $=5.9\cdot10^8L_{\odot}/$ Mpc$^3$  is just 1.4
times greater than the global one, $4.28\cdot10^8L_{\odot}/$ Mpc$^3$
(Jones et al. 2006). Consequently, despite the presence of the
Local Void, the LV represents an overdensity of moderate
amplitude.

A similar relation between the mean density of hydrogen mass
and the radius of sphere, within which it was determined is
shown in the middle panel of Fig.~\ref{Fig20}. The local density
$0.54\cdot10^8L_{\odot}/$ Mpc$^3$ within $D=10$ Mpc is close to
the global mean density of $(0.59\pm0.05)\cdot10^8L_{\odot}/$
Mpc$^3$ according to HIPASS (Zwaan et al. 2003) and ALFALFA
(Martin et al., 2010). The top panel of Fig.~\ref{Fig20}
reproduces the variation of the average density of star formation
rate within a fixed distance $D$. The values of $j_{SFR}$
$[M_{\odot} \cdot {\rm yr}^{-1}{\rm Mpc}^{-3}]$, obtained from
$H\alpha$-fluxes of galaxies are shown by circles, and the
estimates made by FUV-fluxes from the GALEX are shown by
triangles.
Agreement between the independent estimates of $j_{SFR}$ can
be considered quite fair, taking into account the uncertainties
related with the corrections for extinction in the FUV-band.
The mean value from $H\alpha$ and FUV-fluxes within 10 Mpc
amounts to $j_{SFR} =(0.014\pm0.003)$. Given a certain
incompleteness of our $H\alpha$ survey on the outskirts of the LV, as well
as incompleteness of the UV-survey in the region of strong
Galactic extinction, this value is consistent with the global
average value  $(0.018\pm0.003) M_{\odot} \cdot {\rm
yr}^{-1}{\rm Mpc}^{-3}$ according to Salim et al. (2007) and James
et al. (2008).

Therefore, the sample of galaxies in the LV is quite a
suitable representative of the Local Universe in plenty of its
characteristics. It should be emphasized, however, that only 40\%
of  galaxies in this sample have their distance estimates more
accurate than (10--15)\%. The painstaking observational task of
measuring the distances to several hundreds of galaxies within 10
Mpc is a rather pressing issue in the cosmology of nearby
universe. Strictly speaking, the catalog presented here should be
called merely a sample of candidate members of the LV.

\acknowledgments
We thank Valentina Karachentseva and Brent Tully for useful comments
and suggestions to improve the text. We are grateful to the anonymous
referee for a prompt report that helped us improve the manuscript.
This work was supported by the Russian Foundation for Basic Research, grants
10--02--00123,
11--02--00639,
11--02--90449
and RFBR-DFG grant 12-02-91338.
We acknowledge the support of the Ministry of Education and Science of
the Russian Federation, grant No 14.740.11.0901 and the project
2012-1.5-12-000-1011-004.
Support for proposals GO 12546, 12877, 12878 was provided by NASA
through grants from the Space Telescope Science Institute, which is
operated by the Association of Universities for Research in Astronomy,
Inc., under NASA contract NAS5--26555.
We acknowledge the usage of the HyperLeda database
(\url{http://leda.univ-lyon1.fr}).
This research has made use of the NASA/IPAC Extragalactic Database
(NED) which is operated by the Jet Propulsion Laboratory, California
Institute of Technology, under contract with the National Aeronautics
and Space Administration.
We also have made extensive use of the following Web services of surveys:
the Galaxy Evolution Explorer (GALEX),
the Two Micron All Sky Survey (2MASS),
the HI Parkes All Sky Survey (HIPASS),
the Arecibo Legacy Fast ALFA Survey (ALFALFA),
the Extragalactic Distance Database (EDD),
the Sloan Digital Sky Survey (SDSS),
the Digitized Sky Surveys (DSS),
the SAO/NASA Astrophysics Data System (ADS).

{}
	 
%\placetable{Tabl1}
%\placetable{Tabl2}

%\input{Table1.tex}
\clearpage
\pagestyle{empty}
\topmargin=+4cm
\hoffset=-2.0cm
\textwidth=18cm
\small

%\tabcolsep
\begin{deluxetable}{lcrccrrrrcrrrllrrrl}
\rotate
\tablecolumns{18}
\tablewidth{0pc}
\tablecaption{Catalog of nearby galaxies}
\tablehead{
\multicolumn{1}{c}{Name}&
\multicolumn{1}{c}{RA J2000.0 Dec.}&
\multicolumn{1}{c}{$a_{26}$}&
\multicolumn{1}{c}{$b/a$}&
\multicolumn{1}{c}{$A_B^G$}&
\multicolumn{1}{c}{$m_{FUV}$}&
\multicolumn{1}{c}{$B_T$}&
\multicolumn{1}{c}{$m_{H\alpha}$}&
\multicolumn{2}{c}{$K_s$}&
%\multicolumn{1}{c}{$K_b$}&
\multicolumn{1}{c}{$m_{21}$}&
\multicolumn{1}{c}{$W_{50}$}&
\multicolumn{1}{c}{$T$}&
\multicolumn{2}{c}{$T_{dw}$}&
%\multicolumn{1}{c}{tsb}&
\multicolumn{1}{c}{$V_h$}&
\multicolumn{1}{c}{$D$}&
\multicolumn{1}{c}{method}\\
\colhead{1}&
\colhead{2}&
\colhead{3}&
\colhead{4}&
\colhead{5}&
\colhead{6}&
\colhead{7}&
\colhead{8}&
\multicolumn{2}{c}{9}&
\colhead{10}&
\colhead{11}&
\colhead{12}&
\colhead{13}&
\colhead{14}&
\colhead{15}&
\colhead{16}&
\colhead{17}}
\startdata
UGC12894          & 000022.5$+$392944 &   1.02 & 0.87 & 0.47 &  17.57 & 16.80 &  19.91 & 14.02 & *  &  15.66  &  34 & 10 & Ir     & L    &  335 &  8.47 & TF     \\
 WLM               & 000158.1$-$152740 &  11.48 & 0.35 & 0.16 &  12.80 & 11.03 &  15.56 &  9.00 &    &  11.19  &  53 &  9 & Im     & N    & $-$122 &  0.97 & TRGB  \\
 And XVIII         & 000214.5$+$450520 &   1.40 & 0.99 & 0.45 & $>23.00$ & 17.00 &        & 12.49 & *  &         &     & $-$3 & Sph    & L    & $-$332 &  1.36 & TRGB  \\
 ESO409$-$015        & 000531.8$-$280553 &   1.20 & 0.46 & 0.07 &  16.08 & 15.15 &  17.21 & 12.74 & *  &  15.28  &  53 &  9 & Im     & N    &  726 &  7.70 & TF  \\
 AGC748778         & 000634.4$+$153039 &   0.32 & 0.52 & 0.28 &  20.28 & 18.90 &        & 16.29 & *  &  18.21  &  16 & 10 & Ir     & L    &  258 &  5.40 & h'  \\
 And XX            & 000730.7$+$350756 &   0.80 & 0.70 & 0.25 &  21.84 & 19.00 &        & 14.67 & *  &         &     & $-$3 & Sph    & X    &      &  0.80 & TRGB  \\
UGC00064          & 000744.0$+$405232 &   1.66 & 0.56 & 0.34 &  16.59 & 15.50 &        & 12.84 & *  &  14.30  &  60 & 10 & Ir     & N    &  305 &  9.60 & TF  \\
ESO349$-$031        & 000813.3$-$343442 &   1.23 & 0.82 & 0.05 &  17.14 & 15.71 &  21.43 & 13.02 &    &  15.53  &  30 & 10 & Ir     & L    &  221 &  3.21 & TRGB  \\
NGC0024           & 000956.4$-$245748 &   7.24 & 0.22 & 0.08 &  14.16 & 12.38 &  16.03 &  8.96 &    &  13.54  & 213 &  5 &        &      &  550 &  9.90 & TF  \\
NGC0045           & 001403.9$-$231056 &   8.51 & 0.69 & 0.09 &  12.69 & 11.55 &  14.41 &  9.09 &    &  11.98  & 172 &  8 &        &      &  465 &  9.20 & TF  \\
\hline
\multicolumn{18}{l}{\small{\textbf{Note.} Only a portion of this table is shown here to demonstrate its form and content. Machine-readable versions of the full table are available.}}\\
\enddata
\label{t:lv}
\end{deluxetable}

%\input{Table2.tex}
%\clearpage
\pagestyle{empty}
\topmargin=+4cm
\hoffset=-2.0cm
\textwidth=15cm
\small

\begin{deluxetable}{lcrrrrrrrrrrrlrr}
\rotate
\tablecolumns{16}
\tablewidth{0pc}
\tablecaption{Global parameters of the nearby galaxies}
\tablehead{
\multicolumn{1}{c}{Name}&
\multicolumn{1}{c}{RA J2000.0 Dec.}&
\multicolumn{1}{c}{$A_{26}$}&
\multicolumn{1}{c}{$i$}&
\multicolumn{1}{c}{$V_m$}&
\multicolumn{1}{c}{$A_B^i$}&
\multicolumn{1}{c}{$M_B^c$}&
\multicolumn{1}{c}{$SB_B$}&
\multicolumn{1}{c}{$\log L_K$}&
\multicolumn{1}{c}{$\log M_{26}$}&
\multicolumn{1}{c}{$\log M_{HI}$}&
\multicolumn{1}{c}{$V_{LG}$}&
\multicolumn{1}{c}{$\Theta_1$}&
\multicolumn{1}{c}{$MD$}&
\multicolumn{1}{c}{$\Theta_5$}&
\multicolumn{1}{c}{$\Theta_j$} \\
\colhead{1}&
\colhead{2}&
\colhead{3}&
\colhead{4}&
\colhead{5}&
\colhead{6}&
\colhead{7}&
\colhead{8}&
\colhead{9}&
\colhead{10}&
\colhead{11}&
\colhead{12}&
\colhead{13}&
\colhead{14}&
\colhead{15}&
\colhead{16}}
\startdata
UGC12894          & 000022.5$+$392944 &  2.78 & 33 &  21 & 0.00 & $-$13.3 & 25.2 &  7.58 &   8.17 &  7.92  &  619 & $-$1.5 & NGC7640         & $-$1.2 &        \\
 WLM               & 000158.1$-$152740 &  3.21 & 90 &  22 & 0.00 & $-$14.1 & 24.8 &  7.69 &   8.27 &  7.83  &  $-$16 &  0.0 & MESSIER031      &  0.2 &    1.74 \\
 And XVIII         & 000214.5$+$450520 &  0.63 & 10 &     & 0.00 &  $-$9.1 & 26.2 &  6.60 &        &        &  $-$44 &  0.4 & MESSIER031      &  0.5 &    1.52 \\
 ESO409$-$015        & 000531.8$-$280553 &  2.65 & 78 &  23 & 0.00 & $-$14.4 & 24.1 &  7.99 &   8.20 &  7.99  &  769 & $-$1.7 & NGC0253         & $-$1.2 &  \\       
 AGC748778         & 000634.4$+$153039 &  0.53 & 70 &   3 & 0.00 & $-$10.0 & 24.9 &  6.27 &   5.64 &  6.51  &  486 & $-$1.6 & NGC0253         & $-$1.3 &         \\
 And XX            & 000730.7$+$350756 &  0.20 & 58 &     & 0.00 &  $-$5.8 & 27.0 &  5.26 &        &        & $-$182 &  2.4 & MESSIER031      &  2.4 &    1.74 \\
 UGC00064          & 000744.0$+$405232 &  4.87 & 66 &  28 & 0.00 & $-$14.8 & 25.0 &  8.15 &   8.65 &  8.58  &  587 & $-$1.7 & DDO217          & $-$1.3 &   $-$1.15 \\
 ESO349$-$031        & 000813.3$-$343442 &  1.15 & 39 &  14 & 0.00 & $-$11.9 & 24.7 &  7.12 &   7.41 &  7.13  &  230 &  0.0 & NGC0253         &  0.1 &    0.51 \\
 NGC0024           & 000956.4$-$245748 & 19.98 & 85 &  95 & 0.64 & $-$18.3 & 24.5 &  9.75 &  10.32 &  8.91  &  606 & $-$1.0 & NGC0045         & $-$0.8 &    0.42 \\
 NGC0045           & 001403.9$-$231056 & 22.88 & 47 & 103 & 0.17 & $-$18.5 & 24.6 &  9.61 &  10.45 &  9.47  &  528 & $-$0.9 & NGC0024         & $-$0.8 &    0.51 \\
\hline
\multicolumn{16}{l}{\small{\textbf{Note.} Only a portion of this table is shown here to demonstrate its form and content. Machine-readable versions of the full table are available.}}\\
\enddata
\label{t:param}
\end{deluxetable}

\topmargin=0cm
\hoffset=0cm

\appendix

\section{List of apparent magnitudes}

\begin{table}[!h]
\caption{List of apparent magnitudes}
\begin{tabular}{lrrcl}
\hline
 Name             & value  &  error &passband  & bibcode\\
 1                & 2      &  3     & 4        & 5      \\
\hline
 UGC12894          &  17.57 &  0.10 & FUV      & 2012LV.Magnitude..K  \\
 UGC12894          &  16.80 &       & B        & 2011LV.Magnitude..K  \\
 UGC12894          &  19.91 &  0.08 & Ha       & 2008ApJS..178..247K  \\
 UGC12894          &  15.66 &       & HI       & 2011LV.HI.........K  \\
 WLM               &  12.80 &  0.05 & FUV      & 2011ApJS..192....6L  \\
 WLM               &  11.03 &       & B        & 1991RC3.9.C...0000d  \\
 WLM               &  15.56 &  0.32 & Ha       & 2008ApJS..178..247K  \\
 WLM               &  9.00  &  0.50 & Ks       & 2012LV.Magnitude..K  \\
 WLM               &  11.19 &       & HI       & 2009ApJ...696..385G  \\
 \hline
 \multicolumn{5}{l}{\small{\textbf{Note.} Only a portion of this table is shown here to demonstrate its form and content.}}\\
\multicolumn{5}{l}{\small{     Machine-readable versions of the full table are available.}}\\
\label{t:mag}
\end{tabular}
\end{table}

The compilation of apparent magnitudes of galaxies is presented in the Table~\ref{t:mag}.
They are collected from the literature and from surveys and are not
corrected for Galactic extinction or for any other effect. The columns
contain the following information:

1. Main name of a galaxy. It corresponds to the main name in the Table~\ref{t:lv}.

2. Apparent total magnitude of a galaxy or its lower limit (marked by
symbol `$<$').
$H\alpha$ and HI fluxes were transformed to magnitudes using
following equations:
$m_{H\alpha}=-2.5\log(F_{H\alpha}) -13.64$ and
$m_{21}=17.4-2.5\log F_{HI}$,
correspondingly.

3. Measurement error. We estimate uncertainty of eyeball $B$-band
measurements, which were made by our team, to be $\sim0.5^m$.

4. Photometric passband. We have collected the data for FUV, B,
$H\alpha$, Ks and HI passbands.

5. Bibcode for a measurement. The measurements of our team are
marked by bibcode 2011LV.Magnitude..K or 2012LV.Magnitude..K.

\section{List of heliocentric velocities}

\begin{table}[!h]
\caption{List of heliocentric velocities}
\begin{tabular}{lrrl}
\hline
 Name             & value  &  error & bibcode\\
 1                & 2      &  3     & 4      \\
\hline
 UGC12894          & $ 335$ &    22 & 1992ApJS...81....5S \\
 WLM               & $-122$ &     2 & 2004AJ....128...16K \\
 And XVIII         & $-332$ &     3 & 2012ApJ...752...45T \\
 ESO409--015       & $ 726$ &    18 & 2005MNRAS.361...34D \\
 AGC748778         & $ 258$ &     2 & 2011AJ....142..170H \\
 And XX            & $-456$ &     3 & 2012arXiv1211.2638W \\
 UGC00064          & $ 305$ &    17 & 2003A\&A...406..829B \\
 ESO349--031       & $ 221$ &     6 & 2004AJ....128...16K \\
 NGC0024           & $ 550$ &    18 & 2005MNRAS.361...34D \\
 NGC0045           & $ 465$ &    18 & 2005MNRAS.361...34D \\
 \hline
 \multicolumn{4}{l}{\small{\textbf{Note.} Only a portion of this table is shown here to demonstrate its form and content.}}\\
\multicolumn{4}{l}{\small{     Machine-readable versions of the full table are available.}}\\
\label{t:vh}
\end{tabular}
\end{table}

The Table~\ref{t:vh} presents the compilation of heliocentric velocities of galaxies.
The catalog is organized as follows:

1. Main name of a galaxy. It corresponds to the main name in the Table~\ref{t:lv}.

2. Heliocentric velocity ($cz$) in km\,s$^{-1}$.

3. The corresponding error.

4. Bibcode for a measurement.

\section{List of inner kinematic}

\begin{table}[!h]
\caption{List of inner kinematic}
\begin{tabular}{lrrl}
\hline
 Name             & value  &  error & bibcode\\
 1                & 2      &  3     & 4      \\
\hline
 UGC12894          &  34 &       & 1992ApJS...81....5S \\
 WLM               &  53 &    18 & 2004AJ....128...16K \\
 ESO409--015       &  53 &       & 2005MNRAS.361...34D \\
 AGC748778         &  16 &     3 & 2011AJ....142..170H \\
 UGC00064          &  60 &       & 2003A\&A...406..829B \\
 ESO349--031       &  30 &    18 & 2004AJ....128...16K \\
 NGC0024           & 213 &       & 2005MNRAS.361...34D \\
 NGC0045           & 172 &       & 2005MNRAS.361...34D \\
 NGC0055           & 169 &    18 & 2004AJ....128...16K \\
 NGC0059           &  50 &     2 & 2000A\&AS..141..469H \\
 \hline
 \multicolumn{4}{l}{\small{\textbf{Note.} Only a portion of this table is shown here to demonstrate its form and content.}}\\
\multicolumn{4}{l}{\small{     Machine-readable versions of the full table are available.}}\\
\label{t:w}
\end{tabular}
\end{table}

The compilation of data on inner kinematics of galaxies is presented
in the Table~\ref{t:w}.

1. Main name of a galaxy. It corresponds to the main name in the Table~\ref{t:lv}.

2. Observed HI line width at $50\%$ level in km\,s$^{-1}$.

3. The corresponding error.

4. Bibcode for a measurement.

\section{List of distances}

\begin{table}[!h]
\caption{List of distances}
\begin{tabular}{lrrcl}
\hline
 Name             & value  &  error &method  & bibcode\\
 1                & 2      &  3     & 4        & 5      \\
\hline
 UGC12894          &  29.64 &       & TF     & 2012LV.Distance...K \\
 WLM               &  24.93 &  0.04 & TRGB   & 2007ApJ...661..815R \\
 And XVIII         &  25.66 &  0.13 & TRGB   & 2008ApJ...688.1009M \\
 ESO409--015       &  29.43 &       & TF     & 2011LV.Distance...K \\
 AGC748778         &  28.66 &       & h'     & 2011ApJ...739L..22C \\
 And XX            &  24.52 &  0.49 & TRGB   & 2008ApJ...688.1009M \\
 UGC00064          &  29.91 &       & TF     & 2011LV.Distance...K \\
 ESO349--031       &  27.53 &  0.18 & TRGB   & 2006AJ....131.1361K \\
 NGC0024           &  29.98 &       & TF     & 2011LV.Distance...K \\
 NGC0045           &  29.82 &       & TF     & 2011LV.Distance...K \\
 \hline
 \multicolumn{5}{l}{\small{\textbf{Note.} Only a portion of this table is shown here to demonstrate its form and content.}}\\
\multicolumn{5}{l}{\small{     Machine-readable versions of the full table are available.}}\\
\label{t:dist}
\end{tabular}
\end{table}

The Table~\ref{t:dist} contains the compilation of distances to nearby galaxies.

1. Main name of a galaxy. It corresponds to the main name in the Table~\ref{t:lv}.

2. Distance modulus.

3. The corresponding error.

4. Method of the estimation.  
(TRGB) --- by the tip of the red giant branch; 
(Cep) --- from the Cepheid luminosity; 
(SN) --- from the Supernova luminosity;
(SBF) --- from galaxy surface brightness fluctuations; 
(mem) --- from galaxy membership in known groups with measured
distances of other members; 
(TF, FP) --- by the Tully-Fisher relation or by the fundamental plane; 
(BS) --- by luminosity of the brightest stars;
(CMD) --- by the color magnitude diagram using  some prominent features
or simultaneous distance and stellar population fitting;
(HB) --- by the horizontal branch;
(RR) --- from the luminosity of RR Lyrae stars;
(PNLF) --- by the planetary nebula luminosity function;
(h, h$^{\prime}$) --- by the Hubble velocity--distance relation at   $H_0$= 73 km s$^{-1}$Mpc$^{-1}$, not
accounting for (h) or in view of (h$^{\prime}$) a certain
Virgocentric flow model; 
(txt) --- by texture of objects indicating their likely proximity.

5. Bibcode for a measurement.

\section{List of bibliographic references}

\begin{table}[!h]
\caption{List of bibliographic references}
\begin{tabular}{ll}
\hline
 Bibcode             & Reference\\
 1                   & 2        \\
\hline
 2013AJ............G & Giovanelli, R. et al., 2013, AJ, submitted\\
 2012SDSS9.C...0000: & SDSS DR9,2012\\
 2012MNRAS.426..665R & Roychowdhury, Sambit et al., 2012, MNRAS, 426, 665\\
 2012MNRAS.425.2083K & Kaisin, S. S. et al., 2012, MNRAS, 425, 2083\\
 2012MNRAS.422.3208S & Sanchez-Gallego, J. R. et al., 2012, MNRAS, 422, 3208\\
 2012MNRAS.420.2924K & Kirby, Emma M. et al., 2012, MNRAS, 420, 2924\\
 2012MNRAS.419.1362Y & Yang, S.-C. et al., 2012, MNRAS, 419, 1362\\
 2012LV.Magnitude..K & Karachentsev, I. D., Measurements of the LVG team, 2012, LVG\\
 2012LV.HI.........K & Karachentsev, I. D., Measurements of the LVG team, 2012, LVG\\
 2012LV.Distance...K & Karachentsev, I. D., Measurements of the LVG team, 2012, LVG\\
\hline
\multicolumn{2}{l}{\small{\textbf{Note.} Only a portion of this table is shown here to demonstrate its form and content.}}\\
\multicolumn{2}{l}{\small{     Machine-readable versions of the full table are available.}}\\
\label{t:bib}
\end{tabular}
\end{table}

The Table~\ref{t:bib} contain the list of references which were used in our work.

1. Bibcode as in Tables 3--6.

2. Bibliographic references.


\begin{thebibliography}{99}
\bibitem{a00} Abazajian K.N., Adelman-McCarthy J.K., Agueros M.A. et al., 2009, ApJS, 182, 543
\bibitem{a01} Bell E.F., McIntosh D.H., Katz N., Weinberg M.D., 2003, ApJS, 149, 289
\bibitem{a02} Belokurov V., Zucker D.B., Evans N.W. et al., 2006, ApJ, 647, L111
\bibitem{a03} Binney J., Merrifield M., 1998, Galactic astronomy, Binney, J. \& Merrifield, M., ed.
\bibitem{a04} Blanton, M.R., Hogg, D.W., Brinkmann, J. et al., 2003, ApJ, 592, 819
\bibitem{a05} Bremnes, T., Binggeli, B., \& Prugniel, P., 2000, A\&AS, 141, 211
\bibitem{a06} Bremnes, T., Binggeli, B., \& Prugniel, P., 1999, A\&AS, 137, 337
\bibitem{a07} Bremnes, T., Binggeli, B., \& Prugniel, P. 1998, A\&AS, 129, 313
\bibitem{a08} Buzzoni A., 2005, MNRAS, 361, 725
\bibitem{a09} Chiboucas K., Karachentsev, I.D., Tully R.B., 2009, AJ, 137, 3009
\bibitem{a11} Cole S., et al., 2001, MNRAS, 326, 255
\bibitem{a12} C\^{o}t\'{e}, S., Freeman, K., Carignian, C., \& Quinn, P.J., 1997, AJ, 114, 1313
\bibitem{a13} Dalcanton J.J., Williams B.F., Seth A.C. et al., 2009, ApJS, 183, 67
\bibitem{a14} de Vaucouleurs G., de Vaucouleurs A., Corwin H., Buta R.J., Paturel G., \&  Fouqu\'{e} P., 1991, Third Reference Catalogue of Bright Galaxies, New-York - Springer-Verlag
\bibitem{a15} Fingerhut R.L., McCall M.L., Argote M. et al., 2010, ApJ, 716, 792
\bibitem{a16} Fouqu\'{e} P., Paturel G., 1985, A\&A, 150, 192
\bibitem{a17} Fukugita M., Peebles P.J.E., 2004, ApJ, 616, 643
\bibitem{a18} Fukugita M., Shimasaku K., Ichikawa T., 1995, PASP, 107, 945
\bibitem{a19} Gil de Paz A., et al., 2007, ApJS, 173, 185
\bibitem{a21} Giovanelli R., Haynes M.P., Kent B.R. et al., 2005, AJ, 130, 2598
\bibitem{a22} Haynes M.P., Giovanelli R., Martin A.M. et al., 2011, AJ, 142, 170
\bibitem{a23} Huchtmeier, W.K., Karachentsev, I.D., \& Karachentseva, V.E., 2003, A\&A, 401, 483
\bibitem{a24} Huchtmeier, W.K., Karachentsev, I.D., \& Karachentseva, V.E., 2001, A\&A, 377, 801
\bibitem{a25} Huchtmeier, W.K., Karachentsev, I.D., Karachentseva, V.E. \& Ehle M., 2000, A\&AS, 141, 469
\bibitem{a26} Ibata R., Martin N.F., Irwin M. et al., 2007, ApJ, 671, 1591
\bibitem{a27} James P.A., Knapen J.H., Shane N.S. et al., 2008, A\&A, 482, 507
\bibitem{a28} Jarrett, T.N., Chester, T., Cutri R. et al., 2000, AJ, 119, 2498
\bibitem{a29} Jarrett T.H., Chester T., Cutri R., Schneider S. E., Huchra J. P., 2003, AJ, 125, 525
\bibitem{a31} Jones D.H., Peterson B.A., Colless M., Saunders,W., 2006, MNRAS, 369, 25
\bibitem{a32} Kaisin S.S., Karachentsev, I.D., 2008, A\&A, 479, 603
\bibitem{a33} Kaisin S.S., Karachentsev, I.D., 2006, Astrofizika, 49, 337
\bibitem{a34} Kaisina E.I., Makarov D.I., Karachentsev, I.D., Kaisin S.S., 2012, AstBu, 67, 115
\bibitem{a35} Karachentsev, I.D., Nasonova O.G., Courtois H.M., 2011, ApJ, 743, 123
\bibitem{a36} Karachentsev, I.D., Kaisin S.S., 2010, AJ, 140, 1241
\bibitem{a37} Karachentsev, I.D., Kaisin S.S., 2007, AJ, 133, 1883
\bibitem{a38} Karachentsev, I.D., Kashibadze O.G., Makarov, D.I., Tully R.B., 2009, MNRAS, 393, 1265
\bibitem{a39} Karachentsev, I.D., 2005, AJ, 129, 178
\bibitem{a41} Karachentsev, I.D., Karachentseva, V.E., Huchtmeier, W.K., Makarov, D.I., 2004, AJ, 127, 2031 (CNG)
\bibitem{a42} Karachentsev I.D., Makarov D.I., 1999, in Procceedings of IAU Symp. No 186, Kyoto, 109
\bibitem{a43} Karachentsev I.D., Makarov D.I., 1996, AJ, 111, 794
\bibitem{a44} Karachentsev I.D., Makarov D.I., \& Huchtmeier W.K., 1999, A\&AS, 139, 97
\bibitem{a45} Karachentsev I.D., 1994, Astron. Astrophys. Trans., 6, 1
\bibitem{a46} Karachentseva V.E., \& Karachentsev,I.D., 2000, A\&AS, 146, 359
\bibitem{a47} Karachentseva V.E., Karachentsev,I.D., \& Richter G.M., 1999, A\&AS, 135, 221
\bibitem{a48} Karachentseva,V.E., \& Karachentsev I.D., 1998, A\&AS, 127, 409 
\bibitem{a49} Kennicutt R.C., Lee J.C., Funes J.G. et al., 2008, ApJS, 178, 247
\bibitem{a51} Kennicutt R.C., 1998, ARA\&A, 36, 189
\bibitem{a52} Kilborn V.A., Webster R.L., Staveley-Smith L. et al., 2002, AJ, 124, 690
\bibitem{a53} Koribalski B.S., Staveley-Smith L., Kilborn V.A. et al., 2004, AJ, 128, 16
\bibitem{a54} Kova\^{c} K., Oosterloo T.A., van der Hulst J.M., 2009, MNRAS, 400,743
\bibitem{a55} Kraan-Korteweg R.C., 1986, A\&AS, 66, 255
\bibitem{a56} Kraan-Korteweg, R.C. \& Tammann, G.A., 1979, AN, 300, 181
\bibitem{a57} Lee J.C., Gil de Paz A., Kennicutt R.C., et al., 2011, ApJS, 192
\bibitem{a58} Lee, M.G., Freedman, W.L., \& Madore, B.F., 1993, AJ, 106, 964
\bibitem{a59} Liske J., Lemon D.J., Driver S.P. et al., 2003, MNRAS, 344, 397
\bibitem{a61} Makarova L.N., Karachentsev I.D., Rizzi L. et al., 2009, MNRAS, 397, 1672
\bibitem{a62} Martin A.M., Papastergis E., Giovanelli R. et al., 2010, ApJ, 723, 1359
\bibitem{a63} Martin N.F., McConnachie A.W., Irwin M. et al., 2009, ApJ, 705, 758
\bibitem{a64} Martin D.C. et al., 2005, ApJ, 619, L1
\bibitem{a65} Masters K.L., 2005, ``Galaxy flows in and around the Local Supercluster'', PhD, Cornell Univ.
\bibitem{a66} Mathews L.D., Gallagher J.S., Littleton J.E., 1995, AJ, 110, 581
\bibitem{a67} Meyer M.J., Zwaan M.A., Webster R.L. et al., 2004, MNRAS, 350, 1195
\bibitem{a68} Navarro J.F., Frenk C.S., White S.D.M., 1996, ApJ, 462, 563
\bibitem{a69} Papai P. \& Szapudi I., 2010, ApJ, 725, 2078
\bibitem{a71} Pasquali A., Larsen S., Ferreras I. et al., 2005, AJ, 129, 148
\bibitem{a72} Paturel G., Andernach H., Bottinelli L. et al., 1997, A\&AS, 124, 109
\bibitem{a73} Paturel G., Bottinelli L.,  Fouqu\'{e} P., \& Gouguenheim L., 1996, Catalogue of Principal Galaxies: PGC-ROM, Observatoire de Lyon (PGC)
\bibitem{a74} Peebles P.J.E., Nusser, A., 2010, Nature, 465, 565
\bibitem{a75} Peebles P.J.E., Phelps S.D., Shaya E.J., Tully R.B., 2001, ApJ, 554, 104
\bibitem{a76} Peebles P.J.E., 1993, Principles of Physical Cosmology, Princeton, University Press
\bibitem{a77} Rizzi L., Tully R.B., Makarov D.I. et al., 2007, ApJ, 661, 815
\bibitem{a78} Roberts, M.S., \& Haynes, M.P., 1994, Annu. Rev. Astron. Astrophys., 32, 115
\bibitem{a79} Salim S., Rich R.M., Charlot S. et al., 2007, ApJS, 173, 267
\bibitem{a81} Sandage A., \& Tammann G.A., Revised Shapley-Ames Catalog of Bright Galaxies, Carnegie Inst. of Washington, Publ. 635, 1981
\bibitem{a82} Schlegel D.J., Finkbeiner D.P., \& Davis M., 1998, ApJ, 500, 525
\bibitem{a83} Spergel D.N. et al. 2007, ApJS, 170, 377
\bibitem{a84} Staveley-Smith L., Juraszek S., Koribalski B.S. et al., 1998, AJ, 116, 2717
\bibitem{a85} Tonry J.L., Stubbs C.W., Lykke K.R. et al., 2012, ApJ, 750, 99
\bibitem{a86} Tully R.B., Shaya E.J., Karachentsev I.D. et al., 2008, ApJ, 676, 184
\bibitem{a87} Tully R.B., 1988, Nearby Galaxies Catalog, Cambridge University Press
\bibitem{a88} Tully, R.B., \& Fouqu\'{e} P., 1985, ApJS, 58, 67
\bibitem{a89} Vaduvescu O., Richer M.G., McCall M.L., 2006, AJ, 131, 1318
\bibitem{a91} Vaduvescu O., McCall M.L., Richer M.G., Fingerhut R.L., 2005, AJ, 130, 1593
\bibitem{a92} Verheijen, M.A.W., 2001, ApJ, 563, 694
\bibitem{a93} Weisz D.R., Dalcanton J.J., Williams B.F. et al., 2011, ApJ, 739, 5
\bibitem{a94} Willman B., Dalcanton J.J., Martinez-Delgado D. et al., 2005, ApJ, 626, L85
\bibitem{a95} Wong O.I., Ryan-Weber E.V., Garcia-Appadoo et al., 2006, MNRAS, 371, 1855
\bibitem{a96} Zwaan M.A., Staveley-Smith L., Koribalski B.S. et al., 2003, AJ, 125, 2842
\end{thebibliography}
\end{document}